\documentclass[10pt,3p,review,authoryear]{elsarticle}
\journal{Elsevier Journal}

\usepackage[dvipsnames]{xcolor}
\usepackage{color} 
\usepackage{soul,framed} 
\usepackage{mathtools}
\usepackage{lscape}
\usepackage{multirow}
\colorlet{shadecolor}{yellow}
\usepackage{natbib}
\usepackage[ruled,vlined]{algorithm2e}
\usepackage{amsmath}
\usepackage{amsfonts}
\usepackage{amssymb}
\usepackage{subcaption}
\usepackage{hyperref}
\usepackage{lscape}
\usepackage{booktabs}
\usepackage{longtable}
\usepackage{array}
\usepackage[utf8]{inputenc}
\usepackage{placeins}
\usepackage{blindtext}
\usepackage{dirtytalk}
\usepackage{adjustbox}
\usepackage{tabularx}
\usepackage{pdfpages} 
\usepackage{pgffor}
\usepackage{xr}
\makeatletter
\AtBeginDocument{\let\LS@rot\@undefined}
\cleardoublepage\makeatother

\usepackage[shortlabels]{enumitem}
\usepackage{mdwtab}
\usepackage{eqparbox}
\usepackage{url}
\newtheorem{definition}{Definition}

\usepackage{mathtools}
\usepackage{dsfont}
\usepackage{bbold}
\usepackage[english]{babel} 
\usepackage{blindtext}
\usepackage{epstopdf}
\DeclareGraphicsExtensions{.png,.pdf,.jpg}
\usepackage{float}
\usepackage{hyperref}
\hypersetup{
    colorlinks=true,
    linkcolor=blue,
    filecolor=magenta,      
    urlcolor=cyan,
    pdftitle={Overleaf Example},
    pdfpagemode=FullScreen
    }   
\usepackage[utf8]{inputenc}
\usepackage{blindtext}

\makeatletter
\def\ps@pprintTitle{%
  \let\@oddhead\@empty
  \let\@evenhead\@empty
  \let\@oddfoot\@empty
  \let\@evenfoot\@empty
}
\makeatother

\begin{document}

\begin{frontmatter}

\title{
Probabilistic Forecasting of Climate Policy Uncertainty: The Role of Macro-financial Variables and Google Search Data
}


\author{Donia Besher$^1$, Anirban Sengupta$^2$, Tanujit Chakraborty$^{1,3}$\\ 
{\scriptsize \textsuperscript{1} SAFIR, Sorbonne University Abu Dhabi, United Arab Emirates.}\\
{\scriptsize \textsuperscript{2} Indian Institute of Management, Bodhgaya, India.}\\
{\scriptsize \textsuperscript{3} Sorbonne Center for Artificial Intelligence, Sorbonne University, Paris, France.}}


\begin{abstract}
Accurately forecasting Climate Policy Uncertainty (CPU) is essential for designing climate strategies that balance economic growth with environmental objectives. Elevated CPU levels can delay regulatory implementation, hinder investment in green technologies, and amplify public resistance to policy reforms, particularly during periods of economic stress. \textcolor{black}{Despite the growing literature documenting the economic relevance of CPU, forecasting its evolution and understanding the role of macro-financial drivers in shaping its fluctuations have not been explored. This study addresses this gap by presenting the first effort to forecast CPU and identify its key drivers.} We employ various statistical tools to identify macro-financial exogenous drivers, alongside Google search data to capture early public attention to climate policy. Local projection impulse response analysis quantifies the dynamic effects of these variables, revealing that household financial vulnerability, housing market activity, business confidence, credit conditions, and financial market sentiment exert the most substantial impacts. \textcolor{black}{These predictors are incorporated into a Bayesian Structural Time Series (BSTS) framework to produce probabilistic forecasts for both US and Global CPU indices.} Extensive experiments and statistical validation demonstrate that BSTS with time-invariant regression coefficients achieves superior forecasting performance. We demonstrate that this performance stems from its variable selection mechanism, which identifies exogenous predictors that are empirically significant and theoretically grounded, as confirmed by the feature importance analysis. \textcolor{black}{From a policy perspective, the findings underscore the importance of adaptive climate policies that remain effective across shifting economic conditions while supporting long-term environmental and growth objectives.} By providing probabilistic forecasts of CPU, this study enables policymakers to conduct informed risk assessment, anticipate periods of heightened uncertainty, and design strategies that sustain climate objectives under varying economic conditions.
\end{abstract}

\end{frontmatter}

\section{Introduction}
Since the early $19\text{th}$ century, industrialization has served as the principal catalyst of economic growth across both developed and developing nations; by the close of the $20\text{th}$ century, this role had increasingly shifted to the service sector, which emerged as the dominant force driving global economic expansion \cite{grossman1995economic, chevallier2011evaluating}. However, this economic expansion has been accompanied by severe environmental degradation, particularly due to the emission of greenhouse gases (GHGs) such as carbon dioxide ($CO_2$) from industries and automobiles \cite{shahbaz2019foreign, sobrino2014impact, dogan2016co, balsalobre2022influence}. In response, the Kyoto Protocol\footnote{For more details, refer to the official UNFCCC page on the Kyoto Protocol: \href{https://unfccc.int/kyoto_protocol}{https://unfccc.int/kyoto\_protocol}.} was adopted on December 11, 1997 to mitigate this concern. The Kyoto Protocol required developed economies to reduce their emissions by an average of 5.2\% below 1990 levels during the commitment period. However, the Paris Agreement\footnote{The Paris Agreement, adopted at COP21 in Paris, aimed to prevent the average global temperature from increasing by more than 2 degrees Celsius above preindustrial levels and to pursue efforts to limit the increase to 1.5 degrees Celsius. For more details, refer to the official UNFCCC page on the Paris Agreement: \href{https://unfccc.int/process-and-meetings/the-paris-agreement}{https://unfccc.int/process-and-meetings/the-paris-agreement}.}, signed in 2015, expanded the responsibility to both the developing and developed economies (\cite{frayer2021india}). This broader commitment was also incorporated into the Sustainable Development Goals\footnote{For more details, refer to the official UN page on the Sustainable Development Goals: \href{https://sdgs.un.org/goals}{https://sdgs.un.org/goals}.} (SDGs), proposed by the United Nations in 2015, which call for global cooperation to balance environmental sustainability with economic progress. In particular, SDG 13 (Climate Action) and SDG 8 (Decent Work and Economic Growth) reflect the dual challenge of reducing emissions while sustaining development. 

A central challenge in achieving sustainable growth lies in managing the uncertainty surrounding climate-related policy decisions. \textcolor{black}{To quantify this uncertainty, Climate Policy Uncertainty (CPU) indices have been constructed for the United States (US) and globally \cite{gavriilidis2021measuring, ma2024coupling}. The US CPU index captures fluctuations in uncertainty associated with domestic legislative and regulatory developments, while the Global CPU index reflects uncertainty stemming from international climate policy developments and multilateral regulatory actions.} Forecasting CPU is particularly relevant in light of the complex trade-offs that governments face between economic recovery and environmental commitments. Elevated levels of uncertainty lead to hesitation in long-term investments, especially in sectors sensitive to environmental regulation, such as energy, finance, and manufacturing \cite{liang2022, xu2022}. Such uncertainty can delay the transition toward cleaner technologies and hinder progress toward emission reduction goals \cite{yang2024, raza2023, berestycki2022measuring}. Given the global need for substantial investments in low-carbon infrastructure, CPU represents a significant deterrent to both private and public sector decision-making \cite{liang2022, berestycki2022measuring}. Moreover, CPU interacts with economic cycles and public perception. During recessions, policymakers often prioritize expansionary fiscal and monetary policies to stimulate consumption \cite{halkos2016effects, wu2024examining}, potentially sidelining climate mitigation policies \cite{heutel2012should, ide2020recession, annicchiarico2022business, sultanuzzaman2024exploring}. Public attention also shifts toward economic recovery, often viewing climate change as an obstacle \cite{drews2016explains, ge2021impact}. For example, during the post-COVID recovery, fiscal stimulus increased $CO_2$ emissions in China and MINT (Mexico, Indonesia, Nigeria, and Turkey) countries, as production rebounded from 2020 lows \cite{adebayo2023role}. This trend is similar to the increase in $CO_2$ emissions in the US during 2022, making this trade-off a politically sensitive concern \cite{DANISMAN2025123760}. Therefore, understanding the extent to which the state of the economy drives changes in CPU can help policymakers design more resilient environmental regulations.

Economic indicators provide a broad picture of the state of the economy and underpin government and central bank policies. Gross domestic product (GDP) growth depends on industrial development, which results in employment and boosts national income, but at the expense of increased carbon emissions and threats to global warming due to reliance on non-renewable fossil-based sources of energy \cite{newell2021,PEI2022130974}. This trade-off also extends to unemployment, highlighting the challenges of balancing development with climate goals \cite{Ng2023_UnemploymentCleanEnergyOECD}. In addition, interest rates and inflation play an essential role in the formation of climate change policies \cite{akan2024}. Expansionary monetary policies stimulate credit flows thereby increasing global liquidity and driving industrial activity, which further exacerbates global warming risks, whereas inflation discourages the adoption of cleaner, advanced technologies for industrial purposes. These interactions highlight how macroeconomic dynamics can amplify or mitigate environmental risks, making it essential to examine them within key economies that drive global trends.

The US, as the world’s largest economy and the second-largest emitter of GHGs, plays a pivotal role in shaping global climate governance \cite{michaelowa2015rapidly, wang2019decoupling}. Its economic cycles and financial market dynamics set the direction for international capital flows and policy responses \cite{miranda2020us}. Historical patterns reveal that GHG emissions in the US closely track economic activity. \textcolor{black}{The $CO_2$ emissions fell sharply during the COVID-19 pandemic\footnote{US Environmental Protection Agency: \href{https://www.epa.gov/ghgemissions/sources-greenhouse-gas-emissions}{https://www.epa.gov/ghgemissions/sources-greenhouse-gas-emissions}.}. However, $CO_2$ emissions rebounded cumulatively by roughly 8\% in 2022 relative to the 2020 pandemic low due to continued fossil fuel combustion \cite{epa2024ghg}. Between 2020 and 2021, $CO_2$ emissions rose by 7\%, followed by a more modest 1\% increase between 2021 and 2022, reflecting the economic reopening and subsequent recovery.} This pattern suggests that US business and financial cycles are deeply intertwined with climate outcomes and policy responses, underscoring the relevance of forecasting CPU. Supporting this view, \cite{GAIES2025125746} demonstrated that financial stress in the US significantly affects China’s CPU, implying that short-run macroeconomic conditions in the US can propagate CPU internationally. Meanwhile, the Environmental Kuznets Curve theory states that economic growth and human capital development encourage the adoption of cleaner technologies \cite{kuznets1955economic}. These considerations collectively motivate an investigation into how US macroeconomic and financial variables contribute to forecasting CPU.

\textcolor{black}{Despite the growing recognition of CPU as a critical determinant of investment behavior, financial stability, and environmental outcomes, the existing literature has primarily examined its consequences rather than its predictability.} Empirical studies have focused on the effects of CPU on volatility in renewable energy markets \cite{xu2022, liang2022}, corporate investment decisions \cite{matzner2024firms}, and sovereign credit risk \cite{naifar2024spillover, AROURI2025124229, yousaf2025effect}, consistently documenting that elevated CPU levels amplify financial and economic risks. This absence of research on the reverse relationship motivates our analysis. \textcolor{black}{Consequently, this study constitutes the first systematic attempt to forecast CPU.} Building on the above discussion regarding the interplay between economic activity, financial conditions, public perception, and policy uncertainty, we forecast CPU using a comprehensive set of 137 variables covering a broad range of macroeconomic and financial cycle indicators in the US, complemented by Google Trends data, which captures real-time shifts in public concern related to climate policy. \textcolor{black}{Our analysis considers two US CPU indices to capture domestic dynamics and a Global CPU index to evaluate how US economic and financial conditions influence CPU globally.}

Forecasting CPU presents substantial methodological challenges due to the high dimensionality of the covariate space and the presence of structural change and persistent volatility. Most standard econometric and machine learning approaches struggle in this setting, either due to overfitting or a loss of interpretability. To address these challenges, we employ a Bayesian Structural Time Series (BSTS) framework \cite{scott2014} to forecast CPU. BSTS is particularly well suited to this setting because it performs variable selection through its spike and slab prior on the regression coefficients, shrinking irrelevant covariates toward zero while retaining the most informative predictors. Moreover, BSTS is well-suited for policy-relevant forecasting as its credible intervals provide an explicit quantification of forecast uncertainty, enabling an assessment of the likelihood and magnitude of future CPU shocks. Such information is crucial for anticipating investment responses to regulatory uncertainty, guiding the timing and scale of climate policy interventions, and supporting more resilient policy design.

\textcolor{black}{The suitability of BSTS for complex and uncertain forecasting environments is well established in the literature. In macroeconomics and finance, both univariate and multivariate extensions of BSTS have been successfully applied to forecasting stock portfolio returns, where the Bayesian framework effectively controls overfitting, captures cross-series dependence, and accommodates cyclical dynamics, yielding superior predictive performance relative to competing linear and multivariate time-series models \cite{Qiu2018}. Related work demonstrates the strengths of dynamic BSTS variants in energy and financial markets. In particular, dynamic BSTS models have been used to forecast crude oil prices using large information sets, enabling the identification of core economic drivers, structural turning points, and major regime shifts such as the global financial crisis \cite{Lu2020}. In broader financial applications, BSTS has been employed to forecast stock prices in highly uncertain environments, where careful specification of state components has been shown to outperform traditional methods such as Holt-Winters and Seasonal Autoregressive Integrated Moving Average (SARIMA) \cite{Katarina2023}. Beyond financial markets, BSTS has also been applied in tourism economics to forecast international tourist arrivals under political instability and economic disruption, consistently delivering lower forecast errors than competing models and exhibiting robustness to regime changes and prolonged downturns \cite{Chaiboonsri2025, Kimpton2023}. These studies highlight the ability of BSTS to integrate high-dimensional predictors, identify economically relevant variables, and remain robust to nonstationarity and evolving market conditions. These features make it particularly well-suited for forecasting CPU and providing policy-relevant insights into how macroeconomic conditions, financial cycles, and public attention shape uncertainty surrounding climate policy decisions.
}

\textcolor{black}{The main contribution of this paper is that, to the best of our knowledge, it represents the first attempt to forecast CPU. While prior research examines the effects of CPU on macroeconomic and financial markets, the question of whether CPU can be forecasted has remained unexplored. By reversing the prevailing direction of analysis in the literature, this study shifts the focus from the consequences of CPU to its underlying macro-financial predictors, thereby opening a new line of inquiry with direct relevance for environmental policy design and economic planning. Building on this main contribution, we introduce a macro-financial perspective to CPU forecasting by theoretically motivating how business cycle conditions and financial dynamics influence CPU and empirically validating these mechanisms. We identify a relevant subset of predictors using four complementary screening techniques and quantify the dynamic effect of unexpected changes in these variables on CPU via local projection impulse responses. We also incorporate Google search data to demonstrate that public attention to climate policy adds predictive power beyond traditional macro-financial variables. Finally, we empirically show that BSTS produces robust probabilistic forecasts and identifies economically meaningful predictors through analysis of the feature importance plot. Our analysis provides actionable insights into the dynamics of CPU, offering guidance for designing adaptive and resilient climate policies and supporting informed environmental and economic decision-making.}


The remainder of the paper is structured as follows. \textcolor{black}{Section~\ref{sec:lit_review} provides a literature review motivating forecasting with macroeconomic and financial cycle variables.} Section~\ref{sec:data} describes the datasets, summarizes their statistical properties, and highlights how CPU differs from general Economic Policy Uncertainty (EPU). \textcolor{black}{Section~\ref{sec:drivers} outlines the strategy for identifying CPU determinants and the criteria for incorporating Google search data.} Section~\ref{sec:methodology} introduces the BSTS framework. Section~\ref{sec:results} presents the empirical findings, including statistical significance, insights from impulse response and feature importance analyses, and uncertainty quantification. Section~\ref{sec:discussion} discusses the policy implications of the findings. Finally, Section~\ref{sec:conclusion} concludes the paper and highlights potential directions for future research.

\section{Literature Review}\label{sec:lit_review}
\textcolor{black}{CPU is closely intertwined with macroeconomic and financial conditions. For example, \cite{annicchiarico2022business} and \cite{sultanuzzaman2024exploring} reveal that CPU is linked to the business cycle conditions. During periods of economic expansion, governments are more likely to design and implement environmental policies addressing climate change and sustainability. Favorable fiscal balances, stronger employment, and higher corporate profitability increase the willingness of firms, households, and policymakers to comply with and support such regulations \cite{mildenberger2017tradeoff, meyer2022recession}. By contrast, economic downturns weaken firms’ commitment to invest in the technologies and infrastructure needed for regulatory compliance due to reduced profitability. Recessions also shift governmental priorities toward immediate economic stabilization, often relegating long-term environmental objectives to secondary importance \cite{Kahn2011climate, obani2016impact, Bakaki2018environmental, kenny2020protection}.}

\textcolor{black}{Macroeconomic indicators such as GDP, unemployment, and capacity utilization shape expectations regarding the financial and political feasibility of climate regulation. For instance, low-capacity utilization can prompt firms to defer green investments in order to hedge against the risk of the rollback
of stringent environmental regulations under economic pressure from the government \cite{heutel2016climate, annicchiarico2022business}. Moreover, CPU is influenced by credit and financial market conditions. An increase in debt on balance sheets can demotivate the public and industrial sectors to comply with climate-related regulations \cite{bolton2023debt, barucci2025debt}. Macro-financial factors, including corporate bond yields, credit spreads, bank lending, and system-wide leverage, affect CPU because the implementation of climate policy requires high capital investments, tight credit markets, and rising interest rates, which can have an adverse impact on the adoption of climate
policies. Investment in renewable energy depends on the availability of credit and the cost of capital. Tighter financial conditions, such as an increase in interest rates and term spreads, increase the cost of new investments required for decarbonization and heighten uncertainty about the scale and persistence of regulatory commitments \cite{best2017energy, Lessmann2024climate}. Furthermore, \cite{bolton2023carbon} and \cite{carattini2023climate} demonstrate that transition risk and expectations regarding environmental policy are embedded in equity and credit markets. This implies that fluctuations in financial conditions directly shape perceptions of future regulatory regimes.}

\textcolor{black}{Energy and commodity prices are another structural determinant of CPU. Increases in oil prices raise production and transportation costs, as well as household energy expenditures, leading to higher inflation. This effect heightens the short-term economic cost of climate policies and can reduce public willingness to bear transition costs. \cite{hamilton2009oil, kilian2009oil}, and \cite{bouri2022greenenergy} underline such interaction between energy prices, inflation, and policy uncertainty. Asset market valuations and volatility further influence CPU by reflecting forward-looking expectations about the profitability of carbon-intensive versus green technologies and the credibility of regulatory commitments. Risk aversion tends to increase during periods of high market volatility, leadin to policy ambiguity and uncertainty regarding long-term climate strategies \cite{bloom2009uncertainty, baker2016measuring, Pindyck2007uncertainty,ma2024coupling}. Overall, the literature supports our approach of forecasting CPU using macroeconomic and financial variables, as these factors capture key channels through which economic and financial conditions influence CPU. We further examine these channels empirically via impulse response analysis in Section~\ref{sub:impluse_analysis} to quantify how fluctuations in these variables dynamically affect CPU.}

\section{Data Overview and Descriptive Analysis}\label{sec:data}

This study aims to forecast three monthly CPU indices. Although CPU is conceptually related to the broader Economic Policy Uncertainty (EPU) index \cite{baker2016measuring}, the two indices differ fundamentally in scope and implications. While EPU captures uncertainty surrounding general macroeconomic, fiscal, and monetary policy decisions, CPU specifically reflects uncertainty tied to environmental and climate policies. Evidence shows that CPU provides distinct predictive insights for green energy markets \cite{Hong2024CPUGreen}, corporate investment \cite{huang2025CPU}, green innovation \cite{Bai2023Green}, and the adoption of climate-aligned technologies \cite{Chang2024Investment}. This evidence highlights CPU as a more relevant indicator for assessing uncertainty arising from environmental governance both domestically and globally.

To capture these dynamics, we consider a \textit{primary US CPU index} \cite{gavriilidis2021measuring} that captures monthly fluctuations in climate policy as reported by eight major newspapers: \textit{Wall Street Journal, Boston Globe, Los Angeles Times, Chicago Tribune, New York Times, Tampa Bay Times, Miami Herald,} and \textit{USA Today}. \textcolor{black}{To ensure robustness across construction and coverage, we include an \textit{alternative US CPU index} and a \textit{Global CPU index} \cite{ma2024coupling}. The alternative US index is derived from a comparable news-based textual analysis but differs in sources and aggregation, providing an independent measure of CPU in the US. The Global CPU index is constructed using news articles from major newspapers across twelve countries: China, Japan, Korea, India, South Africa, the US, Canada, Brazil, the United Kingdom, France, Germany, and Australia. Following a keyword-based textual analysis framework, CPU is first measured at the country level and then aggregated into a global index using purchasing power parity (PPP)-adjusted GDP weights. This weighting scheme ensures that larger economies exert proportionate influence on the global measure. The resulting index captures the dynamics of CPU associated with major international climate events and global regulatory developments.} Alongside the CPU indices, we compiled a comprehensive set of macroeconomic and financial cycle indicators capturing aggregate economic activity, credit markets, asset prices, labor markets, and housing dynamics. \textcolor{black}{This setup allows us to assess whether US-based macro-financial predictors remain informative for global CPU. 
} To incorporate behavioral and public sentiment, we also include climate-related search terms derived from Google Trends. The resulting integrated dataset enables analysis of how macro-financial conditions and public attention jointly influence CPU.

The primary US CPU index used in this analysis is a monthly series running from April 1987 to June 2023, comprising 435 observations, and is sourced from the Economic Policy Uncertainty website (\url{https://www.policyuncertainty.com/climate_uncertainty.html}). \textcolor{black}{The alternative US and Global CPU indices are monthly series spanning January 2000 to June 2023, with 282 observations, and are obtained from the China Energy and Environmental Policy Research Center Network website (\url{http://www.cnefn.com}).}  We also investigated 137 variables predominantly from the FRED-MD dataset \cite{mccracken2016fred} depicting the US macroeconomic and financial cycle variables. All of these variables are listed and briefly described in Table~\ref{tab:causal_analyses} in Appendix~\ref{causal_ana_res}. These variables include key macroeconomic indicators such as GDP, inflation, unemployment rates, labor market metrics, credit conditions, and financial indices, providing a comprehensive view of the US economy. \cite{mccracken2016fred} proposed presenting this large macroeconomic dataset quarterly. Later, a monthly version of this dataset was introduced. To ensure transparency and reproducibility, further details about the transformations applied to this data are provided in \cite{mccracken2016fred} and summarized in Table~\ref{tab:financial_cycle_variables}. This dataset has been earlier used to forecast economic policy, commodity risk premiums, and US GDP in \cite{carriero2018measuring, rad2023commodity}; and \cite{moramarco2024financial}.

\begin{table*}[t]
\caption{\textcolor{black}{Financial cycle variables, where $x_t$ denotes the respective time series and $x_{t-k}$ denotes the respective time series with $k$ lags. Source details adapted from \cite{moramarco2024financial}.}}
\centering
\footnotesize
\renewcommand{\arraystretch}{1.3}
\begin{tabularx}{\textwidth}{l l l l X}
\toprule
\textbf{Variable Name} & \textbf{Label} & \textbf{Source} & \textbf{Transformation} & \textbf{Mathematical Construction} \\
\midrule
Business Confidence Index & bci & OECD & $x_t$ & Normalized index measuring business sentiment \\
Composite Leading Indicator & cli & OECD & $x_t$ & Amplitude adjusted leading indicator of economic activity \\
Real S\&P 500 Index Growth & sp500 & Shiller & $\ln(x_{t-1}) - \ln(x_{t-4})$ & 3-month lagged quarterly log growth of S\&P500 \\
Cyclically Adjusted Price Earnings Ratio & cape & Shiller & $\ln(x_t)$ & Ratio of current share price to 10-year inflation adjusted earnings \\
Real Credit Growth & cred & FRED & $\ln(x_{t-1}) - \ln(x_{t-4})$ & 3-month lagged quarterly log growth of real credit \\
Credit-to-GDP Ratio & cred\_gdp & BIS & $x_t$ & Ratio of credit to private non-financial sector to nominal GDP \\
Real House Price Growth & hpi & Shiller & $\ln(x_{t-1}) - \ln(x_{t-4})$ & 3-month lagged quarterly log growth of house prices \\
Cyclically Adjusted Price-to-Rent Ratio & capr & FRED & $\ln(x_t)$ & Ratio of house price index to rent price index \\
Real Mortgage Debt Growth & mortg & FRED & $\ln(x_{t-1}) - \ln(x_{t-4})$ & 3-month lagged quarterly log growth of mortgage debt \\
Household Mortgage-to-Income Ratio & mortg\_inc & FRED & $x_t$ & Ratio of household mortgage debt liability to personal income \\
Residential Investment-to-GDP Ratio & prfi\_gdp & FRED & $x_t$ & Ratio of private residential fixed investment to nominal GDP \\
Interest Payments-to-Income Ratio & pip\_inc & FRED & $x_t$ & Ratio of personal interest payments to personal income \\
National Financial Conditions Index & nfci & FRED & $x_t$ & Chicago Fed National Financial Conditions Index \\
\bottomrule
\end{tabularx}
\label{tab:financial_cycle_variables}
\end{table*}

  

\textcolor{black}{The summary statistics and global characteristics of the training samples for the primary US, alternative US, and Global CPU indices are reported in Tables~\ref{tab:summary_statistics} and~\ref{tab:charac_statistics}, respectively. The training period for the primary US CPU index spans April 1987 to June 2021, while the training samples of the alternative US and Global CPU indices run from January 2000 to June 2021.} Table~\ref{tab:summary_statistics} presents key descriptive measures, including the coefficient of variation (CoV), which measures variability relative to the mean, and entropy, which captures the complexity and uncertainty inherent in the distribution of the CPU index. \textcolor{black}{Across all three indices, the CoV values indicate substantial variability relative to the mean, with the alternative US CPU index exhibiting the highest relative dispersion.} Table~\ref{tab:charac_statistics} reports key time series properties including distributional shape (skewness and kurtosis), nonlinearity, seasonality, and long-range dependence \cite{hyndman2018forecasting}. \textcolor{black}{All three CPU series are right-skewed, with the primary US CPU index exhibiting pronounced asymmetry and heavier tails, while the alternative US and Global CPU indices display more moderate skewness and kurtosis. These distributional characteristics indicate a non-negligible probability of elevated CPU episodes across indices, albeit with varying magnitudes.} The time series plots portrayed in Table~\ref{tab:time_acf_pacf} reveal significant variability in CPU, marked by periods of intensified uncertainty in policy direction. The plots depict a clear upward trend across indices, reflecting an increase in climate-related policies over time. While the trend is generally stable, noticeable spikes do correspond to major world events or changes in climate policy. \textcolor{black}{Linearity tests based on Tsay’s and Keenan’s one-degree procedures indicate nonlinear dynamics for the primary US and Global CPU indices, whereas the alternative US CPU index exhibits predominantly linear behavior. Seasonality is detected exclusively in the primary US CPU index using the Ollech and Webel test, while all three series are found to be nonstationary according to the Kwiatkowski-Phillips-Schmidt-Shin (KPSS) test. Finally, the Hurst exponent provides evidence of the presence of long-range dependence in all three CPU series.} This characteristic is further examined through autocorrelation analysis. The autocorrelation function (ACF) and partial autocorrelation function (PACF) plots in Table~\ref{tab:time_acf_pacf} reveal a pronounced autocorrelation structure, indicating that past values of the CPU series exert a persistent influence on future values across multiple lags. Given the presence of such properties, it is essential to adopt a modeling framework that explicitly accounts for structural components.

\begin{table}[h!]
\caption{\textcolor{black}{Summary statistics of the training samples for the three CPU indices.}}
\centering
\scriptsize
\begin{adjustbox}{width=\textwidth, height = 0.95cm}
\begin{tabular}{ccccccccc}
\hline
\textbf{CPU Index} & \textbf{Min Value} & \textbf{Q1} & \textbf{Median} & \textbf{Mean} & \textbf{Q2} & \textbf{Max Value} & \textbf{CoV} & \textbf{Entropy}\\\hline
Primary (US) & 28.162 & 63.563 & 84.166 & 94.876 & 108.046 & 346.612 & 49.777 & 6.019 \\\hline
Alternative (US) & 0.120 & 0.881 & 1.425 & 1.485 & 1.945 & 4.114 & 56.953 & 5.547 \\\hline
Global & 28.500 & 67.591 & 86.115 & 90.895 & 109.719 & 215.088 & 37.603 & 5.553 \\\hline
\end{tabular}
\label{tab:summary_statistics}
\end{adjustbox}
\end{table}

\begin{table}[h!]
\caption{\textcolor{black}{Statistical characteristics of the training samples for the three CPU indices.}}
\centering
\begin{adjustbox}{width=\textwidth, height = 0.95cm}
\begin{tabular}{ccccccc}
\hline
\textbf{CPU Index} & \textbf{Skewness} & \textbf{Kurtosis} & \textbf{Linearity} & \textbf{Seasonality} & \textbf{Stationarity} & \textbf{Long Range Dependence}\\\hline
Primary (US) & 1.863 & 4.531 & Nonlinear & Seasonal & Nonstationary & 0.780  \\\hline
Alternative (US) & 0.664 & 0.376 & Linear & Nonseasonal & Nonstationary & 0.699  \\\hline
Global & 0.849 & 0.871 & Nonlinear & Nonseasonal & Nonstationary & 0.759  \\\hline
\end{tabular}
\label{tab:charac_statistics}
\end{adjustbox}
\end{table}

        
        
        

\begin{table*}[ht]
\small 
\centering
\caption{\textcolor{black}{Time series of the training samples for the three CPU indices, together with their ACF and PACF plots.}}
\label{PLOTS:ACF}
\begin{adjustbox}{max width=\textwidth}
\begin{tabular}{p{6cm} p{4cm} p{6cm}}
\hline
\hspace{2.1cm}Training data & \hspace{1.9cm}ACF Plot & \hspace{2.9cm}PACF Plot\\ 
\hline

\multicolumn{3}{l}{\hspace {1cm} \textbf{Primary US CPU Index}}\\
\multicolumn{3}{c}{\begin{minipage}{\textwidth}
\includegraphics[width=168mm, height=35mm]{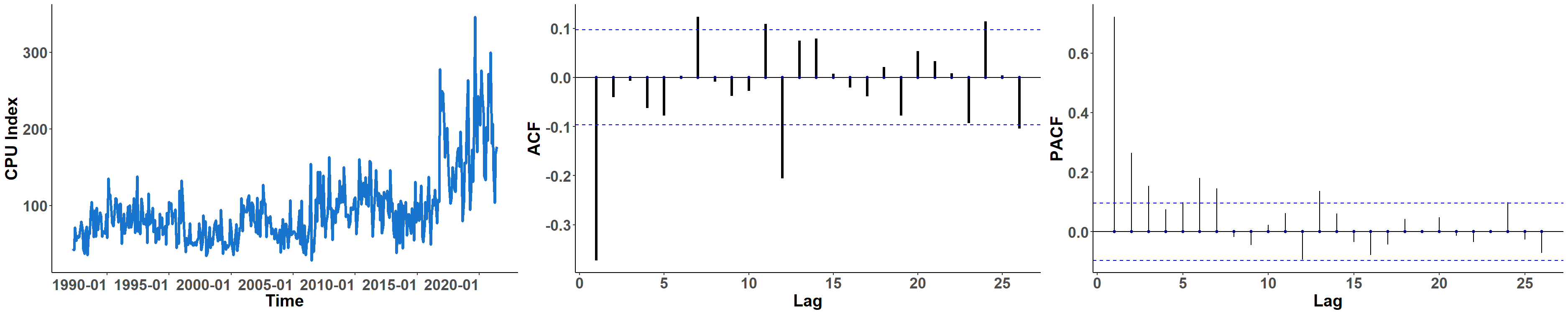}
\end{minipage}}\\
\hline

\multicolumn{3}{l}{\hspace {1cm}\textbf{Alternative US CPU Index}}\\
\multicolumn{3}{c}{\begin{minipage}{\textwidth}
\includegraphics[width=168mm, height=35mm]{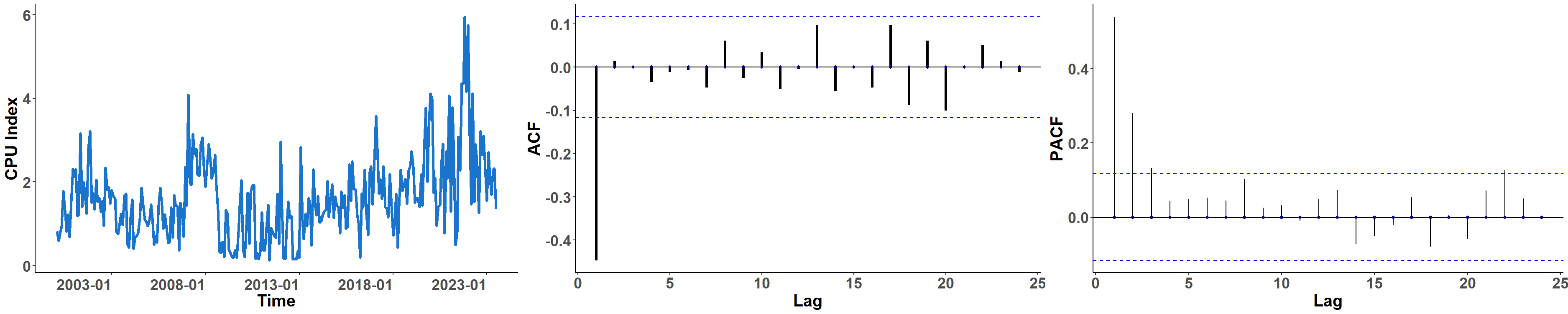}
\end{minipage}}\\
\hline

\multicolumn{3}{l}{\hspace {1.65cm}\textbf{Global CPU Index}}\\
\multicolumn{3}{c}{\begin{minipage}{\textwidth}
\includegraphics[width=168mm, height=35mm]{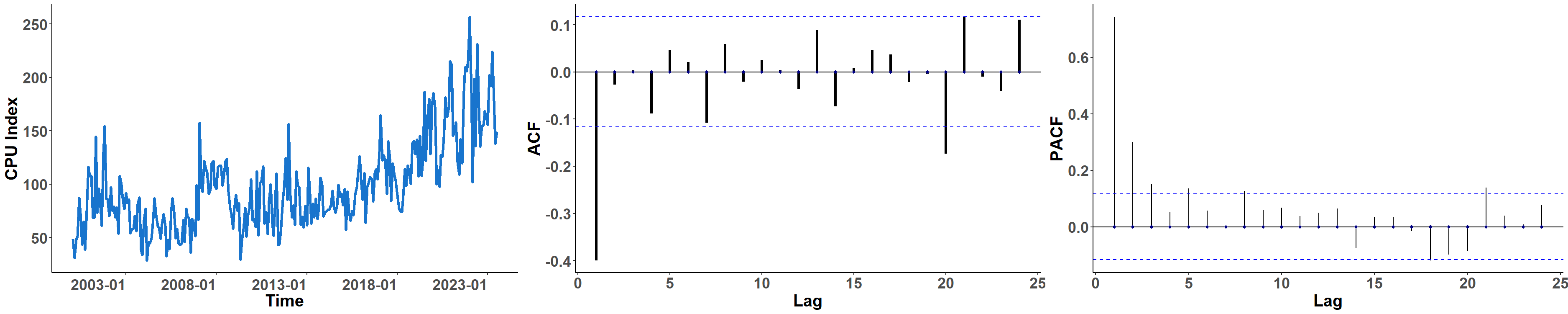}
\end{minipage}}\\
\hline

\end{tabular}
\end{adjustbox}
\label{tab:time_acf_pacf}
\end{table*}






\section{Identifying the Predictors of the CPU Index}\label{sec:drivers}
\textcolor{black}{This section examines the association between macroeconomic, financial, and information-based factors with the CPU index. We begin by describing the empirical screening frameworks used to identify relevant predictors (Section~\ref{sub:variable _selection}), followed by an outline of the theoretical justification for the selected variables (Section~\ref{sub:motivation}).} Finally, we discuss the role of information variables derived from Google search data (Section~\ref{sub:google_trends}).

\subsection{Variable Screening and Predictor Selection}\label{sub:variable _selection}
\textcolor{black}{This section describes the multi-stage screening framework used to identify predictors that are empirically associated with movements in the CPU index.} The objective is to reduce the dimensionality of a large initial set of macroeconomic and financial variables while retaining predictors that exhibit robust and consistent relationships with CPU dynamics. To this end, we apply four complementary screening techniques: Transfer Entropy, Granger Causality, Cross-Correlation analysis, and Wavelet Coherence. Detailed descriptions of these methods and empirical results are provided in Appendix~\ref{Causal_anal}. \textcolor{black}{The screening procedures focus on the primary US CPU index developed by \cite{gavriilidis2021measuring}, which provides a sufficient number of observations to support reliable estimation of nonlinear dependencies, Granger causality, and wavelet coherence. This index therefore serves as the benchmark series for variable selection. Importantly, the temporal dynamics of the primary US and Global CPU indices remain closely aligned, as shown in Table~\ref{tab:time_acf_pacf}, indicating that predictors identified using the US series are expected to capture relevant associations that extend beyond national boundaries. The alternative US CPU index is used as an additional robustness check, as its distinct volatility and persistence properties allow us to assess whether the selected predictors reflect structural drivers of US CPU rather than features tied to a specific index construction. 
}

\textcolor{black}{The four screening tests provide complementary perspectives on the associations between candidate variables and the CPU index. Transfer entropy captures nonlinear directional relationships and information flow, Granger Causality identifies short-run linear predictive dependencies, cross-correlation highlights lead-lag patterns and temporal synchronization, and wavelet coherence uncovers time- and frequency-specific co-movement, revealing how relationships evolve across scales. In terms of implementation, transfer entropy is applied in a pairwise framework as an initial screening device, Granger causality tests are conducted on the differenced series to ensure stationarity, since as the primary US CPU index is shown to be nonstationary in Table~\ref{tab:charac_statistics}, and wavelet coherence is implemented using the ``Morlet" mother wavelet with non-dimensional parameter specific to the wavelet function. Statistical significance is assessed via 100 Monte Carlo randomizations. Phase differences are plotted to capture lead-lag relationships across both time and frequency scales, and a representative example of the resulting coherence plots is shown in Fig.~\ref{fig:Wavelet} in Appendix~\ref{causal_ana_res}.
}

\textcolor{black}{Variable retention follows a systematic rule designed to mitigate the effects of highly correlated predictors, reduce redundancy, and limit confounding effects arising from common macroeconomic factors. A variable is considered relevant if it is supported by at least two screening procedures, which reduces sensitivity to spurious associations. When multiple variables convey closely related economic information, only one representative series is retained to avoid redundancy and multicollinearity concerns. In such cases, preference is given to the variable exhibiting stronger and more consistent empirical support, measured by significance across a larger number of screening tests. For example, CES0600000008 (Average Hourly Earnings: Goods-Producing) and CES2000000008 (Average Hourly Earnings: Construction) capture wage dynamics within closely related segments of the goods-producing sector, with construction representing a subset of the broader goods-producing category. Consequently, they are influenced by similar labor market conditions and exhibit strong co-movement, leading to redundant information content. Hence, between CES0600000008 and CES2000000008, only the latter is retained, as it is supported by three screening procedures compared to two for the former. Finally, a very limited number of substantively important variables, such as the real S\&P 500 index growth, are retained despite being supported by only one screening procedure, reflecting their well-established relevance in the macro-financial literature.} This screening process yields a final set of 60 predictors that are used in the subsequent empirical analysis, with complete screening results reported in Appendix~\ref{causal_ana_res}.

\subsection{Economic Relevance of Selected Predictors}\label{sub:motivation}
\textcolor{black}{This section provides a theoretical grounding for the macroeconomic and financial cycle variables included in the analysis, highlighting why they are expected to be systematically associated with movements in the CPU index.} The goal is to ensure that the set of predictors selected through empirical screening is not only statistically robust but also economically meaningful. These variables capture the broader macro-financial environment that shapes policymakers’ ability to implement, delay, or adjust environmental measures. CPU reflects the interplay of economic conditions, financial and credit constraints, and market sentiment, each of which can influence the timing, scope, and credibility of climate-related policies. By outlining these channels, this section establishes a clear economic rationale for the selected predictors, linking theoretical reasoning with the data-driven selection procedure.

Measures such as the business confidence index (BCI) and composite leading indicator (CLI) capture corporate forward-looking expectations about the economy and their readiness for additional investment. High BCI and CLI levels signal optimism about future growth, profitability, and policy stability. In such conditions, firms and investors perceive lower regulatory risk and are better able to anticipate and adapt to environmental policies. Conversely, when business confidence weakens, the economic environment becomes more uncertain, and firms tend to delay investment decisions, anticipating potential changes in policy direction. These results suggest that CPU is dependent on the state of the underlying economy \cite{zhang2020country}. Therefore, the inclusion of these indicators helps quantify the expectation-driven component of CPU, reflecting how economic sentiment conditions policy credibility and timing. Complementing these indicators, financial market variables further capture the real-time assessment of economic and policy risks by households and investors. Indicators such as the real S\&P 500 index growth, cyclically adjusted price-to-earnings (CAPE) ratio, and credit growth represent the valuation and liquidity conditions underlying economic cycles. Rising equity prices and improving valuations generate a positive wealth effect, enhancing investor and household optimism. This optimism, in turn, lowers the perceived risk of stringent or abrupt policy changes, leading to a decline in CPU \cite{cho2024climate}. However, when asset valuations fall or credit conditions tighten, uncertainty about macroeconomic and regulatory prospects intensifies. 

Housing market indicators, specifically total housing starts (HOUST) and new private housing permits in the northeastern US (PERMITNE), serve as forward-looking measures of construction sector vitality and real estate development. Periods of expansion in housing activity indicate confidence in long-term growth and reflect a robust macroeconomic environment conducive to policy experimentation. In such contexts, households and firms exhibit greater adaptability to environmental regulations and are more likely to invest in climate-friendly technologies, such as energy-efficient housing and renewable installations. Conversely, contractions in housing activity, marked by lower permit issuance or construction starts, often coincide with economic downturns, prompting policymakers to delay or soften environmental measures to protect the real estate sector \cite{cho2024climate}. This relationship suggests that construction and housing-related indicators are likely to exert a stronger influence on CPU dynamics than sector-specific production variables such as manufacturing capacity utilization or mining employment. 

The broader set of housing variables, including the household mortgage-to-income ratio, real house price growth (HPI), cyclically adjusted price-to-rent ratio (CAPR), and private residential fixed investment (PRFI), further elucidate the link between wealth dynamics, financial fragility, and policy sentiment. The mortgage-to-income ratio, in particular, reflects the amount of loans related to the financial stress of individual households due to home loan mortgages, which directly affect public sensitivity to policy reforms that could alter disposable income or asset values, thereby elevating CPU. In contrast, periods of stable or growing housing wealth, reflected by rising HPI or CAPR, tend to strengthen public support for climate-related reforms by reducing perceived financial risk. Hence, housing markets serve as transmission channels for both wealth and policy expectations, rendering them particularly sensitive to shifts in regulatory and macroeconomic conditions \cite{obani2016impact, bumann2021determinants}. 

\textcolor{black}{Elevated household leverage imposes a binding political economy constraint on climate policy implementation. When household balance sheets are highly indebted, sensitivity to inflationary pressures increases, particularly those arising from climate policies such as carbon pricing or energy efficiency mandates that raise utility and compliance costs. In such environments, governments may delay implementation or issue ambiguous signals regarding the timing and scope of environmental regulations to prevent public backlash, thereby increasing CPU. Similarly, a high credit-to-GDP ratio suggests that the economy may be at the peak of a credit cycle, where financial institutions face heightened exposure to leverage and ``stranded asset" risks. In such a scenario, policymakers must then balance decarbonization objectives against concerns over financial stability and credit conditions in carbon-intensive sectors, a trade-off that further amplifies uncertainty surrounding future climate policy actions.}

The real personal consumption expenditures (DPCERA3M086SBEA as in Table~\ref{tab:causal_analyses}) provides insight into aggregate demand conditions, which are central to a consumption-driven economy such as that of the US. Strong consumption growth signals periods of economic expansion and rising public confidence, creating political space for stricter climate action. In contrast, subdued consumption reflects demand-side weakness, which often constrains the implementation of climate initiatives perceived as burdensome to households and firms \cite{zhang2022influence}. Similarly, the personal interest payments-to-income ratio reflects household financial fragility and debt servicing pressures. Higher ratios indicate elevated leverage and reduced disposable income, which can heighten sensitivity to potential cost increases associated with carbon pricing or energy reforms. Thus, periods of rising household indebtedness are likely to coincide with heightened CPU, as policymakers face greater constraints in introducing or maintaining stringent climate measures. Finally, the unemployment rate embodies the trade-off between economic stability and environmental ambition. Elevated unemployment diverts public and political priorities toward immediate economic recovery, reducing tolerance for policies that could impose additional costs. This observation aligns with the findings of \cite{le2025climate}, who document that positive carbon policy shocks can depress aggregate output of the economy by 0.7\% and raise inflation by 0.3\%, reinforcing policymakers’ hesitation to advance environmental agendas during labor market stress.

These findings underscore the association between the selected predictors and the evolution of CPU, providing a theoretically informed perspective on their relationship with CPU. Business confidence and leading economic indicators capture expectation-driven uncertainty, while financial, housing, and credit variables trace the transmission of economic and wealth effects into policy credibility. Labor market and consumption measures reveal how economic conditions shape public support for environmental regulation. Overall, the results show that CPU arises from the interaction of economic sentiment, financial stability, and public welfare expectations, underscoring that stabilizing economic expectations can mitigate climate-related policy uncertainty and strengthen long-term climate commitments.

\subsection{Google Trends Indicators}\label{sub:google_trends}
\begin{figure}[h!]
    \centering
    \includegraphics[width=0.9\textwidth]{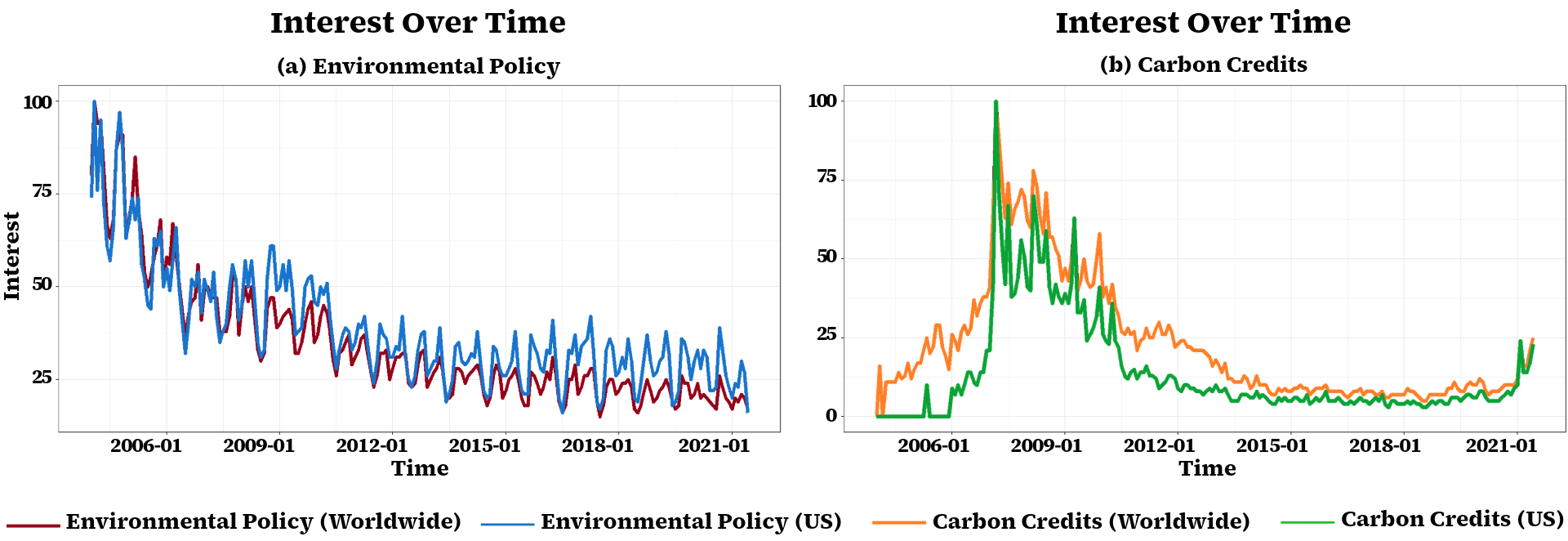}
    \caption{\textcolor{black}{Global and US interest in the Google search terms (a) ``Environmental Policy" and (b) ``Carbon Credits" from January 2004 to June 2021. The series display similar temporal patterns, highlighting the close alignment between US and worldwide search behavior. Data sourced from Google Trends: \url{https://trends.google.com/trends?geo=US&hl=en-GB}.}}
    \label{fig:WW_US}
\end{figure}

Beyond macro-financial predictors, public attention and information diffusion also shape perceptions of climate policy. To capture this behavioral dimension, we incorporate data from ``Google Trends"\footnote{\url{https://trends.google.com/trends?geo=US&hl=en-GB}.} for a broad set of search terms related to climate policy. These include queries such as ``climate policy”, ``climate risk”, ``environmental policy”, ``carbon credits”, and ``clean energy”,  among others.  \textcolor{black}{All search terms are collected at the worldwide level; however, their temporal patterns closely mirror those of US-specific searches, as depicted in Fig.~\ref{fig:WW_US}. The close alignment in both temporal patterns and magnitudes reflects the fact that global Google search intensity is heavily influenced by US policy developments and media coverage, which generate information shocks that dominate worldwide search dynamics. Additionally, global search terms are particularly relevant when forecasting the Global CPU index, which is also analyzed in this study. Consequently, retaining worldwide Google Trends series provides both a reasonable approximation for US behavior and valuable predictive information for CPU.} 

\begin{figure}[h!]
    \centering
    \includegraphics[width=0.9\textwidth]{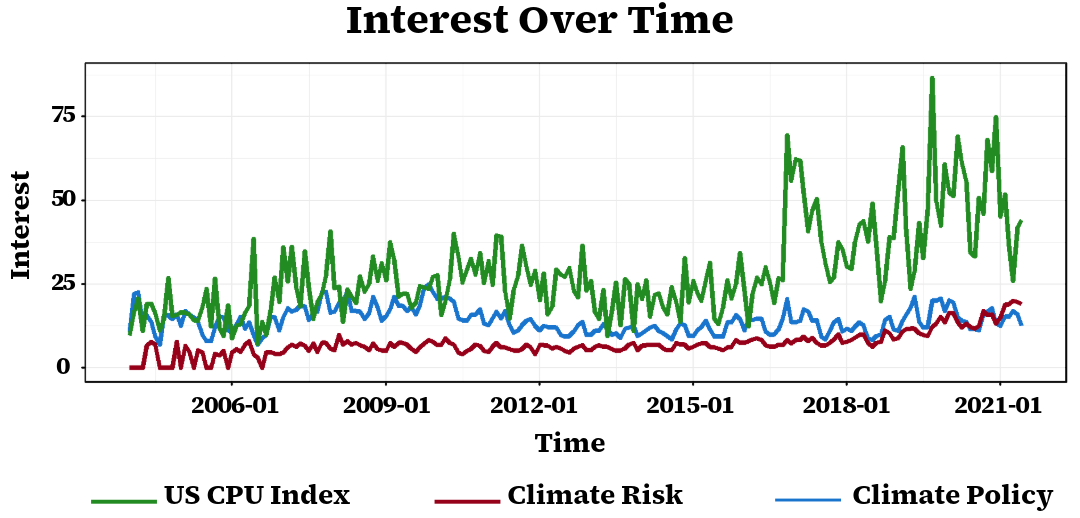}
    \caption{\textcolor{black}{Interest in the worldwide search terms ``Climate Policy" and ``Climate Risk" from January 2004 to June 2021, plotted alongside the primary US CPU index. The three series exhibit similar temporal patterns.}}
    \label{fig:google}
\end{figure}

Incorporating Google Trends data complements macroeconomic and financial indicators by introducing a social attention component, capturing shifts in public concern, awareness, and anticipation of policy initiatives. Fig.~\ref{fig:google} shows clear co-movement between public search behavior and the primary US CPU index, indicating that CPU is influenced not only by economic and policy conditions but also by shifts in collective perception and information flow.  A surge in search activity may correspond to heightened discourse around policy developments or global climate events, such as United Nations climate conferences, which are associated with elevated CPU levels. Conversely, declining search interest can signal stabilization in policy expectations. Although Google Trends provides a valuable proxy for public attention, it can exhibit temporal spikes unrelated to substantive policy changes, introducing potential measurement noise. To address this, we apply the four screening techniques described in Section~\ref{sub:variable _selection} to identify search terms that display robust and consistent co-movement with the primary US CPU index. Only these significant terms are retained for inclusion in the forecasting model, and they are listed in Table~\ref{tab:causal_analyses_google} in Appendix~\ref{causal_ana_res}.

\section{Model Formulation}\label{sec:methodology}
The Bayesian Structural Time Series (BSTS) model, introduced by \cite{scott2014}, provides a flexible approach for forecasting time series data that extends the classical structural time series model by incorporating Bayesian inference to jointly estimate the model parameters and unobserved components. First, the observed series \( y_t \) is decomposed into latent components such as trends, seasonal fluctuations, and the influence of external covariates. Each component is specified within a state-space formulation, which provides a systematic way to model their evolution over time. The Bayesian framework enables the inclusion of prior knowledge, which is beneficial in managing uncertainty about model specification and producing forecasts along with their associated uncertainty through the posterior distribution. This combination makes BSTS particularly useful in economic forecasting problems, where interpretability, rigorous uncertainty assessment, and the integration of diverse economic and sentiment indicators are crucial. 

The BSTS framework relies on the state-space representation, where the observed time series \( y_t \) is associated with an unobserved state vector \( \alpha_t \). This formulation uses the idea that the observable data are generated by latent processes evolving over time. The general form of a state-space model is expressed as:
\begin{equation}
\begin{split}
    y_t &= Z_t^\top \alpha_t + \epsilon_t, \quad \epsilon_t \sim \mathcal{N}(0, H_t) \\
    \alpha_{t+1} &= T_t \alpha_t + R_t \eta_t, \quad \eta_t \sim \mathcal{N}(0, Q_t),
\end{split}
\label{eq}
\end{equation}
where \( y_t \) denotes the observed time series, \( \alpha_t \) represents the latent state vector encompassing the unobserved components of the model. The matrices \( Z_t \) and \( T_t \) are the observation and transition matrices, respectively. The observation matrix \( Z_t \) links the unobserved states to the observed data, while the transition matrix \( T_t \) describes how the unobserved states change over time. The terms \( \epsilon_t \) and \( \eta_t \) are independent and identically distributed Gaussian noise processes with covariance matrices \( H_t \) and \( Q_t \), respectively, reflecting observation and state uncertainties. We consider the observation noise covariance \( H_t \) to be a positive scalar, denoted by \( \sigma_{\epsilon}^2 \), following the specification in \cite{scott2014}. The components of the BSTS framework are detailed in Appendix~\ref{bsts_components}. Together, these components allow the BSTS model to flexibly capture the diverse patterns and dynamics characteristic of macroeconomic time series.

\subsection{Prior Specification}
To handle high-dimensional data, the BSTS model employs the spike and slab prior to induce sparsity by effectively setting most regression coefficients to zero while allowing a subset to vary. This is relevant for CPU forecasting due to the strong association between CPU and a large number of potential macroeconomic and financial cycle variables. This approach begins by introducing an
indicator variable \( \gamma_k \) encoding whether predictor \( x_k \) is included in the model:
\[
\gamma_k = 
\begin{cases} 
1, & \text{if } \beta_k \neq 0 \\ 
0, & \text{if } \beta_k = 0, 
\end{cases}
\]
 where $\beta_k$ is the coefficient of predictor $x_k$. Then, we define \( \beta_{\gamma} \) as the vector of regression coefficients corresponding to predictors with \( \gamma_k = 1 \). The joint prior over the coefficients \( \beta \), indicator vector \( \gamma \), and observation noise variance \( \sigma_\epsilon^2 \), namely the ``spkie and slab'' prior, is formulated as follows:
\[
p(\beta, \gamma, \sigma_\epsilon^2) = p(\beta_{\gamma} | \gamma, \sigma_\epsilon^2) \: p(\sigma_\epsilon^2 | \gamma) \: p(\gamma).
\]
The marginal prior \( p(\gamma) \), referred to as the ``spike'', assigns high probability to zero coefficients for irrelevant predictors, thereby encouraging sparsity. For computational convenience, each \( \gamma_k \) is assumed to follow an independent Bernoulli distribution:
\[
\gamma \sim \prod_{k=1}^{K} \pi_k^{\gamma_k} (1 - \pi_k)^{1 - \gamma_k},
\]
where \( \pi_k \) represents the prior inclusion probability of predictor \( x_k \). In practice, all \( \pi_k \) are set to a common value \( \pi = \frac{\tilde{p}}{K} \), defined as the expected model size, with \( \tilde{p} \) being the expected number of nonzero predictors and \( K \) is the total number of predictors. The ``slab'' corresponds to the conditional conjugate priors on the included coefficients and the variance, \( p(\beta_{\gamma} | \sigma_\epsilon^2, \gamma) \) and \( p\left(\frac{1}{\sigma_\epsilon^2} | \gamma\right) \), respectively. They are modeled as:
\[
\beta_{\gamma} \mid \sigma_\epsilon^2, \gamma \sim \mathcal{N}\left(b_{\gamma}, \, \sigma_\epsilon^2 \left(\Omega_{\gamma}^{-1}\right)^{-1}\right), \quad \frac{1}{\sigma_\epsilon^2} \mid \gamma \sim \text{Gamma}\left(\frac{\nu}{2}, \frac{ss}{2}\right),
\]
\noindent where \( b_{\gamma} \) represents the prior mean vector, \( \Omega_{\gamma}^{-1} \) is a symmetric matrix corresponding to \( \gamma_k = 1 \), and \( ss \) and \( \nu \) denote the expected coefficient of variation (\( R^2 \)) from the regression and the prior sample size, respectively. The precision matrix $\Omega_\gamma^{-1}$ is defined as \small{\( \Omega_\gamma^{-1} = \kappa \: (\omega \: X^\top X + (1 - \omega) \: \text{diag}(X^\top X))/n \)}, where \( X \) is the design matrix, $n$ is the total number of observations, and \( \kappa \) represents the number of significance observations on the prior mean \( b_\gamma = 0\), with \( \omega = \frac{1}{2} \) and \( \kappa = 1 \), following \cite{scott2014}. This formulation enables the model to select the most relevant predictors for the observed time series while excluding irrelevant predictors.

\subsection{Posterior Distribution}
In Bayesian inference, the posterior distribution reflects the updated beliefs about model parameters after incorporating the observed data. This is especially valuable in economic applications, where uncertainty and prior knowledge play a critical role in inference and forecasting. In order to derive the posterior distribution, we first define \( \theta \) as the set of model parameters, excluding the regression coefficients \( \beta \) and the observation noise variance \( \sigma^2_{\epsilon} \). The posterior distribution \( p(\theta \mid y) \) is given by Bayes’ theorem as:
\[
p(\theta \mid y) = \frac{p(y \mid \theta) \, p(\theta)}{p(y)},
\]
where \( p(y \mid \theta) \) is the likelihood of the observed data given the parameters \(\theta\), \( p(\theta) \) is the prior belief about the parameters before observing the data, and \( p(y) \) is the marginal likelihood, acting as a normalizing constant, obtained by integrating out all the parameters. In order to facilitate posterior inference, the observed series is transformed to separate the contributions of the latent state \( \alpha_t \) from those of the covariates by defining \( \mathbf{y_t^*} = y_t - {Z_t^*}^\top \alpha_t \), where \( {Z_t^*}^\top \) is the observation matrix from Eqn.~\eqref{eq} with the regression term \( \beta^\top x_t \) excluded. Given an inclusion vector \( \gamma \), the conditional posterior distributions for the regression coefficients \( \beta_\gamma \) and noise variance \( \sigma^2_{\epsilon} \) are modeled as:
\[
\beta_{\gamma} \mid \sigma_\epsilon^2, \gamma, \mathbf{y^*} \sim \mathcal{N}(\tilde{\beta}_{\gamma}, \sigma_\epsilon^2 (V_{\gamma}^{-1})^{-1}), \quad \frac{1}{\sigma_\epsilon^2} \mid \gamma, \mathbf{y^*} \sim \text{Gamma}\left(\frac{N}{2}, \frac{SS_{\gamma}}{2}\right).
\]
Following the specifications of \cite{scott2014}, the matrix \( V_{\gamma}^{-1} = (X^\top X)_{\gamma} + \Omega_{\gamma}^{-1} \). The posterior mean of the regression coefficients for the selected variables \( \tilde{\beta}_{\gamma} = (V_{\gamma}^{-1})^{-1} \left( X_{\gamma}^\top \mathbf{y^*} + \Omega_{\gamma}^{-1} b_{\gamma} \right) \). The sample size $N$ combines the observed sample size $n$ and the prior sample size $\nu$, i.e., \( N = n + \nu \), and the adjusted sum of squares is given by \( SS_{\gamma} = ss + \mathbf{y^{* \top}} \mathbf{y^*} + b_{\gamma}^\top \Omega_{\gamma}^{-1} b_{\gamma} - \tilde{\beta}_{\gamma}^\top V_{\gamma}^{-1} \tilde{\beta}_{\gamma} \). Intuitively, these expressions update the prior distribution of the regression coefficients \( \beta_\gamma \) by integrating information from the observed data, represented by \( X^\top X \), with prior knowledge encoded in the precision matrix \( \Omega_\gamma^{-1} \) and mean vector \( b_\gamma \). The resulting posterior reflects a weighted balance between the data likelihood and prior beliefs. Subsequently, the marginal posterior distribution for \( \gamma \), which determines which predictors are included, is given by:
\[
\gamma | \mathbf{y}^* \sim C(\mathbf{y}^*) \frac{\left|\Omega^{-1}_{\gamma}\right|^{\frac{1}{2}} p(\gamma)}{\left|V^{-1}_{\gamma}\right|^{\frac{1}{2}} SS_{\gamma}^{\frac{N}{2} - 1}},
\]
where \( C(\mathbf{y^*}) \) is a normalizing constant that is independent of \( \gamma \). The posterior distribution of the BSTS model is difficult to compute explicitly due to its high dimensionality and highly structured dependencies. However, it can be approximated using a Markov Chain Monte Carlo algorithm with the following steps:
\begin{enumerate}
    \item Sample the latent state \( \alpha \) from \( p(\alpha \mid y, \theta, \beta, \sigma_\epsilon^2) \) using the simulation smoother of \cite{9c656354-8aab-357c-856b-281880305042}. This step accounts for the unobserved time-varying components such as trend and seasonality.
    
    \item Sample the parameters \( \theta \) from \( p(\theta \mid y, \alpha, \beta, \sigma_\epsilon^2) \), updating beliefs about the components.

    \item Sample the regression coefficients \( \beta \) and variance \( \sigma_\epsilon^2 \) from a Markov chain with the stationary distribution \( p(\beta, \sigma^2 | y, \alpha, \theta) \).
\end{enumerate}
Setting \( \phi = (\theta, \beta, \sigma_\epsilon^2, \alpha) \), the procedure could be iterated to produce samples \( \phi^{(1)}, \phi^{(2)}, \ldots \), which converge in distribution to the to the true posterior \( p(\phi | y) \) \cite{scott2014}.

\section{Experimental Analysis}\label{sec:results}
\textcolor{black}{This section evaluates the forecasting performance of the BSTS model for multiple measures of CPU, including the primary US CPU index, the alternative US CPU index, and the Global CPU index, and investigates the economic significance of the variables identified by the model.}
We compare the BSTS performance against benchmark statistical, machine learning, and deep learning forecasting models across four time horizons: 3-month-ahead, 6-month-ahead, 12-month-ahead, and 24-month-ahead, to assess its adaptability and robustness. This section is organized as follows. Sections~\ref{baseline} and~\ref{metric} provide concise overviews of the baseline models and the evaluation metrics, respectively. Section~\ref{results} details the experimental results and benchmark comparisons. Section~\ref{stat_sig} discusses the statistical significance of the forecasts. \textcolor{black}{Section~\ref{ablation} provides an ablation study of BSTS models with time-invariant and time-varying coefficients to evaluate the impact of coefficient flexibility on forecasting performance. Section~\ref{sub:impluse_analysis} examines the responses of the CPU index to innovations in key macroeconomic and financial cycle variables, providing empirical support for the theoretical channels discussed in Sections~\ref{sec:lit_review} and~\ref{sub:motivation}.} Section~\ref{feature_plot} complements this analysis by presenting the feature importance plot, highlighting the economic relevance of the variables selected by the BSTS model. Finally, Section~\ref{credible_intervals} presents the credible intervals generated by the BSTS model.

\subsection{Baseline Models}\label{baseline}
The forecasting performance of the BSTS model is evaluated against a range of statistical, machine learning, and deep learning models, all of which can incorporate covariates. The benchmark frameworks include: Autoregressive Integrated Moving Average with exogenous variables (ARIMA-X), Autoregressive Fractionally Integrated Moving Average with exogenous variables (ARFIMA-X), Autoregressive Neural Network with exogenous variables (ARNN-X), Neural Basis Expansion Analysis for Time Series with exogenous variables (NBeats-X), Neural Hierarchical Interpolation for Time Series with exogenous variables (NHiTS-X), Decomposition-based Linear model with exogenous variables (DLinear-X), and Normalization-based Linear model with exogenous variables (NLinear-X). A comprehensive overview of these baseline models can be found in Appendix~\ref{appendix_baseline}.

\subsection{Evaluation Metrics}\label{metric}
Five evaluation metrics are employed to assess the forecasting performance of the competing models. These metrics include Root Mean Squared Error (RMSE), Mean Absolute Error (MAE), Mean Absolute Scaled Error (MASE), Mean Absolute Percentage Error (MAPE), and Symmetric Mean Absolute Percentage Error (SMAPE). These metrics are chosen to provide a comprehensive assessment of forecast accuracy, capturing both absolute and relative errors, sensitivity to large deviations, and scale-independent performance. 
Their mathematical formulations are given by:
 {\scriptsize
\begin{equation*}
 \begin{gathered}
 \text{RMSE} = \sqrt{\frac{1}{h}\sum_{t=1}^{h} (y_t - \hat{y}_t)^2}; \quad \quad \quad
\text{MASE} = \frac{\sum_{t = 1}^{h} |\hat{y}_t - y_t|}{\frac{h}{T-1} \sum_{t = 2}^T |y_t - y_{t-1}|}; \quad \quad \quad
\text{MAE} = \frac{1}{h} \sum_{i=1}^{h} \left| y_i - \hat{y}_i \right|; \\
\text{MAPE} = \frac{1}{h} \sum_{t=1}^h \frac{|\hat{y}_t - y_t|}{|y_t|} \times 100 \%;  \quad \quad \quad
\text{SMAPE} = \frac{1}{h} \sum_{t=1}^h \frac{|\hat{y}_t - y_t|}{\left(|\hat{y}_t|+ |y_t|\right)/2} \times 100\%,
\end{gathered}
\end{equation*}
}
where $y_t$ represents the observed time series, $\hat{y}_t$ is the forecasted value, $h$ refers to the forecast horizon, and $T$ is the length of the in-sample (training) period. According to standard practice, the model with the lowest error metric is regarded as the best-performing model \cite{hyndman2018forecasting}.

\subsection{Experimental Results and Baseline Comparison}\label{results}
To ensure a comprehensive evaluation, the BSTS model is compared against several state-of-the-art forecasters. The implementation of the BSTS model is carried out in \textbf{R} statistical software using the \texttt{bsts} function of the \texttt{bsts} package. The benchmark models ARIMA, ARFIMA, and ARNN are implemented through the \texttt{forecast} package, while the deep learning models NBeats, NHiTS, DLinear, and NLinear are implemented via Python's \texttt{darts} library. \textcolor{black}{Model estimation and forecast evaluation are conducted within a consistent and leakage-free empirical framework. For each forecast horizon $h$, models are trained on all available observations except the last $h$ points, which are reserved as the test set. Hyperparameters are selected using only information available in the training sample. Out-of-sample forecasts are then generated recursively for the test set, ensuring that only information available at time 
$t$ is used in prediction. All predictors are similarly constructed using information available at time $t$, maintaining a consistent real-time forecasting setup. This procedure is applied separately for each forecast horizon, with the test set length and corresponding training window defined by the horizon. The splitting scheme is summarized in Table~\ref{tab:train_test_split}, providing a transparent description of the forecast evaluation design.} The optimal state specifications of the BSTS model, selected by minimizing the RMSE metric, are listed in Table~\ref{BSTS SS}. 

     

\begin{table}[!ht]
\caption{\textcolor{black}{Training and testing periods for forecasting CPU indices across different forecast horizons.}}
\centering
\begin{adjustbox}{width=\textwidth}
\begin{tabular}{llcccc} 
\hline
CPU Index & Horizon & Training Period &  Number of Observations &  Testing Period &  Number of Observations \\ \hline
Primary (US) & $h = 3$  & Apr 1987 - Mar 2023 & 432 & Apr 2023 -- Jun 2023 & 3 \\ 
           & $h = 6$  & Apr 1987 - Dec 2022 & 429 & Jan 2023 - Jun 2023 & 6 \\ 
           & $h = 12$ & Apr 1987 - Jun 2022 & 423 & Jul 2022 - Jun 2023 & 12 \\
           & $h = 24$ & Apr 1987 - Jun 2021 & 411 & Jul 2021 - Jun 2023 & 24 \\ 
\hline
Alternative (US) \& Global & $h = 3$  & Jan 2000 - Mar 2023 & 279 & Apr 2023 - Jun 2023 & 3 \\ 
                         & $h = 6$  & Jan 2000 - Dec 2022 & 276 & Jan 2023 - Jun 2023 & 6 \\ 
                         & $h = 12$ & Jan 2000 - Jun 2022 & 270 & Jul 2022 - Jun 2023 & 12 \\
                         & $h = 24$ & Jan 2000 - Jun 2021 & 258 & Jul 2021 - Jun 2023 & 24 \\ 
\hline
\end{tabular}
\end{adjustbox}    
\label{tab:train_test_split}
\end{table}

\begin{table}[!ht]
\caption{\textcolor{black}{Optimal BSTS state specifications for forecasting the three CPU indices across different forecast horizons. Notation: LL = Local Level, LLT = Local Linear Trend, AR = Autoregressive component, S = Seasonal component.}}
   \centering
   \begin{adjustbox}{width=\textwidth}
    \begin{tabular}{ccccc} \hline
    CPU Index & $(state\;specification)_{h = 3}$ & $(state\;specification)_{h = 6}$& $(state\;specification)_{h = 12}$&  $(state\;specification)_{h = 24}$ \\ \hline
    Primary (US) & LL, AR, S & LLT, AR, S & LLT & LL, LLT, AR, S\\ \hline
    Alternative (US) & LLT & LLT, S & LLT & LL, AR, S\\ \hline
    Global & AR & LL, AR & LL, AR & LLT, AR, S\\ \hline
    \end{tabular}
\label{BSTS SS}
\end{adjustbox}    
\end{table}

Excluding covariates from the forecasting framework undermines the economic interpretability and structural validity of the model. As discussed in Section~\ref{sec:lit_review}, macroeconomic and financial cycle indicators serve as fundamental predictors of the CPU index, capturing key economic linkages and transmission mechanisms that shape climate-related policy uncertainty. Their inclusion enhances the model’s ability to reflect underlying structural dynamics, ensuring that forecasts remain both theoretically consistent and practically informative for economic analysis and decision-making. Therefore, this study focuses on evaluating the additional predictive power gained from incorporating sentiment information. To do this, all models are estimated under two distinct configurations: (i) including the 60 statistically significant macroeconomic and financial cycle indicators (denoted as X$_\text{M}$), and (ii) including both the 60 macro-financial variables together with sentiment-based data derived from Google Trends (denoted as X$_\text{MG}$). This setup enables a systematic comparison of the forecasting performance across both covariate sets.

After implementing BSTS and the baseline models, out-of-sample forecasts were generated for multiple horizons (3-, 6-, 12-, and 24-month-ahead) using the two covariate sets (X$_\text{MG}$ and X$_\text{M}$). \textcolor{black}{Forecasts are generated recursively, with multi-step-ahead predictions obtained sequentially from earlier forecasts, thereby ensuring a realistic assessment of model performance across horizons.} For the primary US CPU index, Tables~\ref{tab:performance_metrics_XMG_PUS} and~\ref{tab:performance_metrics_XM_PUS} report the predictive performance on the test set using the sets X$_\text{MG}$ and X$_\text{M}$, respectively. The results indicate that ARFIMA achieves the highest accuracy for short-term forecasts, corresponding to the 3-month-ahead horizon ($h = 3$), across both covariate sets, followed closely by BSTS. However, BSTS consistently outperforms all baseline models over the medium- and long-term horizons, corresponding to the 6-, 12-, and 24-month-ahead horizons ($h = 6, 12, 24,$ respectively), 
reflecting its ability to capture persistent temporal dependencies that are particularly relevant for climate policy and investment decisions as they typically unfold over extended periods. The superior performance of BSTS can be attributed to its spike-and-slab prior, which effectively identifies the most informative predictors while excluding irrelevant covariates, thereby reducing model complexity, mitigating overfitting, and allowing the model to adapt dynamically to evolving relationships within the data. \textcolor{black}{Importantly, our pre-model variable selection procedure complements this Bayesian regularization by retaining variables with consistent empirical support across multiple screening methods, which mitigates the risk of posterior probability dilution among highly correlated predictors, ensuring that the posterior distribution concentrates on truly relevant predictors.} Notably, the inclusion of Google Trends indicators in X$_\text{MG}$ consistently improves BSTS performance relative to X$_\text{M}$, highlighting the added informational value of behavioral indicators in forecasting CPU.

\textcolor{black}{The performance of the models on the test set for the alternative US CPU index is reported in Appendix~\ref{appendix_emp_res}, Tables~\ref{tab:performance_metrics_XMG_AUS} and~\ref{tab:performance_metrics_XM_AUS}, corresponding to the covariate sets X$_\text{MG}$ and X$_\text{M}$, respectively. Using the set X$_\text{MG}$, BSTS and ARIMA exhibit the strongest accuracy at the 3-month-ahead horizon, depending on the evaluation metric, while ARIMA slightly outperforms BSTS at the 6-month horizon. For the 12-month-ahead horizon, ARNN achieves the highest predictive performance. When using the set X$_\text{MG}$, BSTS and NLinear perform best at the 3-month horizon, whereas ARFIMA delivers superior accuracy at both the 6- and 12-month horizons. Importantly, BSTS demonstrates the strongest performance at the 24-month-ahead horizon across both covariate sets, which remains the most relevant time frame for informing climate policy and decision-making. During the implementation of ARIMA and ARFIMA, a small number of Google Trends indicators were omitted to ensure model convergence. 
}

\textcolor{black}{For the Global CPU index, model performance using the covariate sets X$_\text{MG}$ and X$_\text{M}$ on the test set is reported in Appendix~\ref{appendix_emp_res}, Tables~\ref{tab:performance_metrics_XMG_Global} and~\ref{tab:performance_metrics_XM_Global}, respectively. BSTS attains the highest accuracy for the 3- and 6-month-ahead horizons, while at 12 months its performance is comparable to ARNN, depending on the evaluation metric when using the set X$_\text{MG}$. However, when using the set X$_\text{M}$, short-term accuracy at the 3-month horizon varies across metrics, with BSTS and NLinear performing well, whereas the leading models at the 6- and 12-month horizons differ by metric. Notably, BSTS consistently demonstrates strong predictive performance across both covariate sets at the 24-month-ahead horizon. Hence, the US macroeconomic and financial cycle variables identified from the primary US CPU index encapsulate meaningful and transferable information that is also predictive of Global CPU. This outcome provides empirical support for the variable selection strategy adopted in Section~\ref{sub:variable _selection}. Moreover, it confirms the applicability of BSTS to forecasting CPU indices beyond the US context, indicating that its predictive advantages are not confined to a single national index. While BSTS is capable of handling large sets of covariates, alternative statistical models lack this capacity. Conversely, neural network architectures are black-box models, limiting their applicability in high-stakes decision-making settings where interpretability is desired. 
}

Several notable patterns emerge from the combined results. First, BSTS consistently achieves superior performance at the 24-month-ahead horizon across all indices and covariate sets, underscoring its relevance for long-term climate policy planning. Second, the inclusion of Google Trends indicators markedly enhances BSTS performance, highlighting the value of incorporating public attention and information diffusion into the forecasting framework. Finally, a similar, though less pronounced, improvement is observed for the statistical and machine learning models, suggesting that sentiment-based data can enhance their predictive ability. In contrast, the performance of deep learning models generally declines as the dimensionality of the covariates increases. This limitation is attributed to the absence of an explicit variable selection mechanism, combined with the relatively limited number of observations of the CPU index series, which introduces additional noise and hinders the models’ ability to learn stable patterns. Hence, excluding these additional variables mitigates the risk of overfitting, thus enhancing the accuracy of these models. By comparison, BSTS does not suffer from this degradation, as its spike-and-slab prior enables variable selection and shrinkage, preventing noise accumulation even in high-dimensional covariate spaces. Taken together, these findings underscore both the empirical and policy relevance of the BSTS framework and the central role of macro-financial conditions and sentiment-based Google Trends indicators in shaping the dynamics of CPU.

\begin{table*}[!ht]
\caption{Evaluation of the BSTS-X$_\text{MG}$ model's performance relative to baselines across all forecast horizons for the primary US CPU index using macro-financial variables and Google Trends indicators. The \textbf{\underline{best}} and \textbf{\textit{second-best}} results are highlighted.}
\centering
\scriptsize
\begin{adjustbox}{width=1\textwidth, height= 3.5cm}
\begin{tabular}{cccccccccc}
\hline
Horizon & Metric & ARFIMA-X$_\text{MG}$ & ARIMA-X$_\text{MG}$ & ARNN-X$_\text{MG}$ & BSTS-X$_\text{MG}$ & NBEATS-X$_\text{MG}$ & NHiTS-X$_\text{MG}$ & DLinear-X$_\text{MG}$ & NLinear-X$_\text{MG}$
\\\hline

$h = 3$ & MAPE & \textbf{\underline{2.381}} & 14.147 & 9.368 & \textbf{\textit{4.708}} & 50.601 & 16.053 & 35.628 & 22.305 \\
  & SMAPE & \textbf{\underline{2.329}} & 13.582 & 9.317 & \textbf{\textit{4.598}} & 39.630 & 14.401 & 29.489 & 19.806 \\
  & MAE & \textbf{\underline{5.346}} & 32.034 & 20.996 & \textbf{\textit{10.568}} & 112.928 & 35.616 & 79.831 & 49.823 \\
  & MASE & \textbf{\underline{0.919}} & 5.507 & 3.610 & \textbf{\textit{1.817}} & 19.415 & 6.123 & 13.724 & 8.566 \\
  & RMSE & \textbf{\underline{7.384}} & 38.779 & 21.123 & \textbf{\textit{10.655}} & 118.719 & 42.559 & 87.526 & 53.186 \\\hline
$h = 6$ & MAPE & 16.299 & 10.896 & \textbf{\textit{8.186}} & \textbf{\underline{6.029}} & 55.015 & 31.979 & 33.733 & 24.987 \\
  & SMAPE & 18.155 & 11.424 & \textbf{\textit{8.455}} & \textbf{\underline{6.343}} & 41.896 & 26.388 & 28.258 & 21.895 \\
  & MAE & 40.330 & 26.575 & \textbf{\textit{20.230}} & \textbf{\underline{5.572}} & 129.318 & 72.532 & 77.783 & 57.627 \\
  & MASE & 1.372 & 0.904 & \textbf{\textit{0.688}} & \textbf{\underline{0.530}} & 4.401 & 2.468 & 2.647 & 1.961 \\
  & RMSE & 47.209 & 33.874 & \textbf{\textit{25.702}} & \textbf{\underline{24.201}} & 139.230 & 84.355 & 83.300 & 61.294 \\\hline
$h = 12$ & MAPE & 18.605 & 19.187 & \textbf{\textit{13.795}} & \textbf{\underline{11.345}} & 74.528 & 43.596 & 28.088 & 27.713 \\
  & SMAPE & 21.217 & 21.530 & \textbf{\textit{15.295}} & \textbf{\underline{12.133}} & 52.193 & 34.688 & 25.878 & 25.423 \\
  & MAE & 47.620 & 48.585 & \textbf{\textit{36.038}} & \textbf{\underline{29.361}} & 164.626 & 96.759 & 67.963 & 66.540 \\
  & MASE & 1.365 & 1.393 & \textbf{\textit{1.033}} & \textbf{\underline{0.842}} & 4.719 & 2.774 & 1.948 & 1.907 \\
  & RMSE & 59.830 & 66.708 & \textbf{\textit{59.739}} & \textbf{\underline{44.153}} & 177.004 & 107.107 & 87.748 & 82.833 \\\hline
$h = 24$ & MAPE & 36.951 & 27.282 & \textbf{\textit{24.502}} & \textbf{\underline{18.194}} & 32.408 & 34.603 & 36.579 & 29.985 \\
  & SMAPE & 44.414 & 32.126 & \textbf{\textit{26.489}} & \textbf{\underline{19.182}} & 32.406 & 39.809 & 42.008 & 33.742 \\
  & MAE & 90.807 & 69.591 & \textbf{\textit{59.253}} & \textbf{\underline{44.670}} & 81.390 & 89.026 & 89.146 & 77.392 \\
  & MASE & 1.479 & 1.134 & \textbf{\textit{0.965}} & \textbf{\underline{0.728}} & 1.326 & 1.450 & 1.452 & 1.261 \\
  & RMSE & 103.194 & 91.093 & \textbf{\textit{77.387}} & \textbf{\underline{66.440}} & 109.319 & 116.036 & 112.517 & 102.789 \\\hline
\end{tabular}
\label{tab:performance_metrics_XMG_PUS}
\end{adjustbox}
\end{table*}

\begin{table*}[!ht]
\caption{Evaluation of the BSTS-X$_\text{M}$ model's performance relative to baselines across all forecast horizons for the primary CPU index using macro-financial variables. The \textbf{\underline{best}} and \textbf{\textit{second-best}} results are highlighted.}
\centering
\scriptsize
\begin{adjustbox}{width=\textwidth, height = 3.5cm}
\begin{tabular}{cccccccccc}
\hline
Horizon & Metric & ARFIMA-X$_\text{M}$ & ARIMA-X$_\text{M}$ & ARNN-X$_\text{M}$ & BSTSX$_\text{M}$ & NBEATS-X$_\text{M}$ & NHiTS-X$_\text{M}$ & DLinear-X$_\text{M}$ & NLinear-X$_\text{M}$ \\\hline

h = 3 & MAPE & \textbf{\underline{2.388}} & 10.108 & 8.991 & \textbf{\textit{4.597}} & 39.965 & 21.413 & 6.777 & 5.473 \\
 & SMAPE & \textbf{\underline{2.328}} & 10.732 & 9.498 & \textbf{\textit{4.493}} & 32.744 & 19.306 & 6.576 & 5.586 \\
 & MAE & \textbf{\underline{5.403}} & 22.617 & 20.334 & \textbf{\textit{10.316}} & 89.059 & 47.878 & 15.038 & 12.401 \\
 & MASE & \textbf{\underline{0.929}} & 3.888 & 3.496 & \textbf{\textit{1.774}} & 15.311 & 8.231 & 2.585 & 2.132 \\
 & RMSE & \textbf{\underline{7.983}} & 24.992 & 22.719 & \textbf{\textit{10.384}} & 93.807 & 48.287 & 16.838 & 14.689 \\
\hline
h = 6 & MAPE & 16.299 & 15.348 & \textbf{\textit{8.530}} & \textbf{\underline{6.218}} & 78.979 & 12.845 & 9.379 & 8.825 \\
 & SMAPE & 18.155 & 16.959 & 8.740 & \textbf{\underline{6.514}} & 55.271 & 13.812 & 9.374 & \textbf{\textit{8.526}} \\
 & MAE & 40.330 & 37.013 & \textbf{\textit{17.144}} & \textbf{\underline{15.959}} & 182.182 & 31.902 & 22.284 & 20.990 \\
 & MASE & 1.372 & 1.260 & \textbf{\textit{0.583}} & \textbf{\underline{0.543}} & 6.200 & 1.086 & 0.758 & 0.714 \\
 & RMSE & 47.209 & 41.596 & \textbf{\textit{23.012}} & 24.372 & 190.497 & 43.951 & 24.206 & \textbf{\underline{23.723}} \\
\hline
h = 12 & MAPE & 18.713 & 19.391 & \textbf{\textit{18.233}} & \textbf{\underline{13.109}} & 82.676 & 31.806 & 23.715 & 19.439 \\
 & SMAPE & 21.355 & 22.730 & 21.689 & \textbf{\underline{13.250}} & 57.214 & 27.998 & 22.313 & \textbf{\textit{18.159}} \\
 & MAE & 47.885 & 50.231 & 47.810 & \textbf{\underline{31.904}} & 185.336 & 72.103 & 55.801 & \textbf{\textit{45.274}} \\
 & MASE & 1.373 & 1.440 & 1.371 & \textbf{\underline{0.915}} & 5.313 & 2.067 & 1.600 & \textbf{\textit{1.298}} \\
 & RMSE & 60.100 & 66.534 & 68.297 & \textbf{\underline{41.903}} & 195.770 & 85.485 & 70.668 & \textbf{\textit{56.954}} \\
\hline
h = 24 & MAPE & 36.951 & 30.759 & 29.900 & \textbf{\underline{21.491}} & 31.907 & 47.433 & 26.375 & \textbf{\textit{25.637}} \\
 & SMAPE & 44.414 & 36.677 & 34.549 & \textbf{\underline{23.382}} & 35.416 & 43.404 & \textbf{\textit{28.479}} & 29.233 \\
 & MAE & 90.807 & 77.414 & 72.859 & \textbf{\underline{53.364}} & 81.261 & 113.077 & 67.907 & \textbf{\textit{67.646}} \\
 & MASE & 1.479 & 1.261 & 1.187 & \textbf{\underline{0.869}} & 1.324 & 1.842 & 1.106 & \textbf{\textit{1.102}} \\
 & RMSE & 103.194 & 95.410 & \textbf{\textit{88.810}} & \textbf{\underline{72.952}} & 111.822 & 134.613 & 91.364 & 94.302 \\
\hline
\end{tabular}
\label{tab:performance_metrics_XM_PUS}
\end{adjustbox}
\end{table*}

\subsection{Robustness and Statistical Significance Tests}\label{stat_sig}

\begin{figure}[h!]
 \centering
 \includegraphics[width=0.9\textwidth]{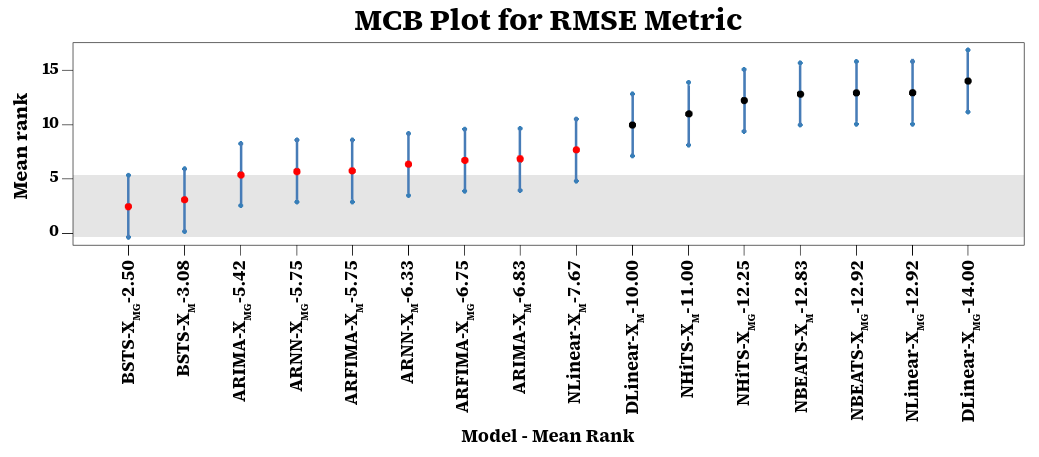}
 \caption{\textcolor{black}{Multiple comparisons with the best (MCB) plot comparing model performance across the two covariate sets and the three CPU indices. BSTS-X$_\text{MG}$-2.50, for instance, signifies that the average ranking of the BSTS-X$_\text{MG}$ model, according to the RMSE metric, is 2.50. This same interpretation holds for other models.}}
\label{fig:MCB_test}
\end{figure}

The forecasting performance of the competing models is evaluated using the model-agnostic Multiple Comparisons with the Best (MCB) procedure to assess the statistical significance of differences in measurement errors. The MCB test is a nonparametric ranking approach that orders each of the $\mathcal{M}$ forecasting models based on their predictive accuracy across $\mathcal{D}$ datasets \cite{koning2005}. The model with the lowest average rank is identified as the best-performing forecasting framework. \textcolor{black}{Fig.~\ref{fig:MCB_test} depicts the ranking of the models across the three CPU indices for the two covariate configurations. The BSTS-X$_\text{MG}$ model achieves the lowest mean rank of 2.50 under the RMSE metric across all forecast horizons, establishing it as the top-performing model, followed by BSTS-X$_\text{M}$ (3.08), and ARIMA-X$_\text{MG}$ (5.42).} The MCB results indicate that BSTS-X$_\text{M}$ performs comparably but does not outperform BSTS-X$_\text{MG}$, underscoring the value of incorporating both macro-financial indicators and Google Trends data when forecasting the CPU index.

The gray-shaded region in Fig.~\ref{fig:MCB_test} represents the upper limit of the critical distance for BSTS-X$_\text{MG}$, serving as the benchmark for comparison. \textcolor{black}{The deep learning models NHiTS, NBEATS, and DLinear exhibit critical intervals well above this threshold, signifying substantially weaker forecasting performance mainly due to the limited size of the training data.} Consistent with the results reported in Section~\ref{results}, models such as ARIMA, ARFIMA, and ARNN demonstrate improved accuracy when the covariate set X$_\text{MG}$ is employed, whereas deep learning architectures tend to deteriorate with the inclusion of additional predictors. This divergence is likely attributed to their increased susceptibility to noise and overfitting in high-dimensional settings, as well as the limited number of observations in some CPU index series, which restricts their capacity to reliably learn complex relationships from a high-dimensional set of exogenous variables. Overall, the MCB analysis confirms statistically significant differences in model performance and provides further evidence that the BSTS framework consistently outperforms alternative models across different forecast horizons.

From Fig.~\ref{fig:MCB_test}, we observe that BSTS-X$_\text{MG}$ achieves higher predictive performance than BSTS-X$_\text{M}$; nonetheless, the difference is marginal. To further demonstrate the substantial explanatory power and forecasting improvement gained from incorporating sentiment-based Google Trends indicators, we employ the Murphy diagram analysis. Unlike the MCB procedure, which ranks models based on their average forecast accuracy, the Murphy diagram evaluates whether a model consistently outperforms competing alternatives across a wide family of scoring functions \cite{ehm2016quantiles, chakraborty2025narfima}. This approach provides a more comprehensive assessment of the robustness of forecast performance. The Murphy diagram is constructed using the following scoring function:
\begin{equation}
\tilde{s}(\hat{y}_t, y_t) = 
    \begin{cases}
    |y_t - \theta|, &  \min(\hat{y}_t, y_t) \leq \theta < \max(\hat{y}_t, y_t) \\
0, & \text{otherwise},
    \end{cases}
    \label{Eq:Theta}
\end{equation}
where \( y_t \) denotes the observed value at time \( t \), and \( \hat{y}_t \) is the forecast generated by the model. The parameter \( \theta \in \mathbb{R} \) determines the shape of the loss function and acts as a threshold sliding between the forecast and the observation, thereby enabling the scoring function to highlight different aspects of forecast errors. Specifically, lower values of \(\theta\) increase the penalty for underestimations, while higher values place greater weight on overestimations. To assess the relative performance of competing forecasting models, the average score for model \( i \) over a forecast horizon of length \( h \) is computed as $\mathcal{S}_i(\theta) = \frac{1}{h} \sum_{t=1}^h \tilde{s}(\hat{y}_{t,i}, y_t)$, where \( \hat{y}_{t,i} \) is the forecast from model \( i \) at time \( t \). Plotting the average scores \( \mathcal{S}_i(\theta) \) across a range of values of \(\theta\) yields the Murphy diagram. Consequently, the Murphy diagram serves as a flexible and insightful tool to identify models that consistently outperform others across diverse scoring functions. To empirically assess the robustness of the BSTS-X$_\text{MG}$ model relative to benchmark methods, we use the \texttt{murphydiagram} package in \textbf{R} to construct Murphy diagrams for all three CPU indices. \textcolor{black}{Murphy diagrams compare BSTS-X$_\text{MG}$ against BSTS-X$_\text{M}$ and ARIMA-X$_\text{MG}$, the strongest competing models according to the RMSE-based MCB test.}

\begin{figure}
    \centering
    \includegraphics[width=0.86\textwidth]{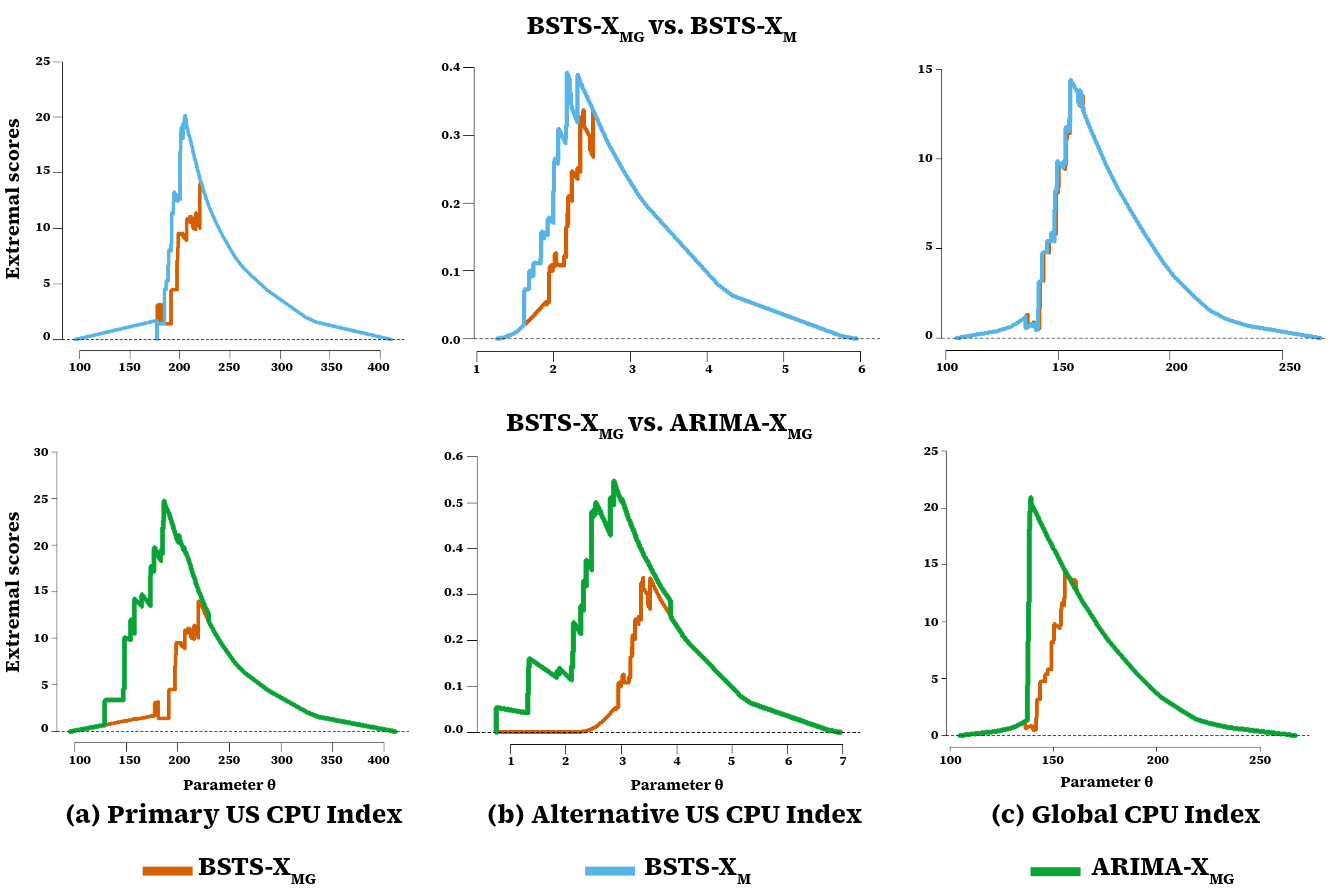}
    \caption{\textcolor{black}{Murphy diagrams of BSTS-X$_\text{MG}$ with baselines (BSTS-X$_\text{M}$ (top) and ARIMA-X$_\text{MG}$ (bottom)) for the 24-month-ahead forecasts for the (a) primary US, (b) alternative US, and (c) Global CPU indices. The parameter $\theta$ represents the shape parameter as defined in Eqn.~\eqref{Eq:Theta}. Lower scores indicate better performance.}}
    \label{fig:MurphyDiagram}
\end{figure}

Fig.~\ref{fig:MurphyDiagram} presents the Murphy diagrams for the 24-month-ahead horizon. Lower extremal scores indicate superior predictive accuracy. \textcolor{black}{The results for both the US CPU indices indicate that BSTS-X$_\text{MG}$ consistently outperforms BSTS-X$_\text{M}$ and ARIMA-X$_\text{MG}$ across a broad range of values for $\theta$, with the performance gap being particularly pronounced for low to mid-range thresholds.} This outcome reinforces the findings of the MCB analysis and underscores the significant forecasting value of integrating sentiment-based Google Trends indicators. \textcolor{black}{In contrast, for the Global CPU index, the performance differential between BSTS-X$_\text{MG}$ and BSTS-X$_\text{M}$ is negligible across the range of $\theta$ values considered. This suggests that, at the global level, Google Trends variables provide little incremental predictive value beyond macro-financial covariates. Fig.~\ref{fig:MurphyDiagram} further demonstrates that BSTS consistently outperforms ARIMA across all CPU indices, providing empirical support for the Bayesian model’s superior long-horizon forecasting ability. The relative advantage of BSTS-X$_\text{MG}$, however, diminishes when $\theta$ exceeds approximately 225 for the primary US CPU index, 4 for the alternative US CPU index, and 160 for the Global CPU index, indicating that performance differences narrow when overprediction penalties dominate.} By combining insights from both the MCB and Murphy analyses, we establish robust evidence of the forecasting superiority of the BSTS framework and the substantial predictive contribution of Google Trends indicators in forecasting CPU indices.

\subsection{Ablation Study: Time-Invariant versus Time-Varying Regression Coefficients}\label{ablation}
\textcolor{black}{The policy-sensitive nature of the CPU index raises the question of whether allowing regression coefficients to evolve over time improves forecasting performance, as assuming time-invariant coefficients in this context implicitly limits the ability to capture structural changes and interpret policy-driven dynamics. To validate this, we conducted an ablation study comparing BSTS models with time-invariant coefficients (BSTS-X$_\text{MG}^\text{TI}$) against models with time-varying coefficients (BSTS-X$_\text{MG}^\text{TV}$), using the optimal state-space specifications reported in Table~\ref{BSTS SS} and the covariate set X$_\text{MG}$, which was identified in Sections~\ref{results} and~\ref{stat_sig} as most predictive. Results for all CPU indices and forecast horizons are reported in Table~\ref{tab:ablation}.}

\textcolor{black}{The results shows that BSTS with time-invariant coefficients consistently outperforms the time-varying specification. For the primary US CPU index, BSTS-X$_\text{MG}^\text{TI}$ achieves superior accuracy across all horizons. For the alternative US and Global CPU indices, BSTS-X$_\text{MG}^\text{TV}$ only surpasses the time-invariant model at the 6- and 12-month horizons, respectively, while BSTS-X$_\text{MG}^\text{TI}$ provides better forecasts for all remaining horizons. The superior performance of the BSTS model with time-invariant coefficients can be attributed to two key factors. First, several CPU index series have a limited number of observations, constraining the ability of BSTS-X$_\text{MG}^\text{TV}$ to learn stable patterns, particularly when combined with a large set of covariates. Second, the covariate set X$_\text{MG}$ already captures the most relevant macro-financial and behavioral information, reducing the need for additional flexibility in the regression coefficients. Hence, fixing the coefficients mitigates the risk of overfitting, ensuring more robust long-horizon predictions. These findings indicate that, for the CPU index, incorporating time-varying regression coefficients is not effective. 
Consequently, the use of time-invariant coefficients in BSTS-X$_\text{MG}$ is empirically justified and supports informed climate policy planning and investment decisions without introducing unnecessary model complexity.}

\begin{table}[ht]
\centering
\caption{\textcolor{black}{Evaluation of  BSTS-X$_\text{MG}$ models with time-invariant and time-varying regression coefficients (BSTS-X$_\text{MG}^\text{TI}$ and BSTS-X$_\text{MG}^\text{TV}$, respectively) across all forecast horizons and CPU indices using macro-financial variables and Google Trends indicators. The \textbf{\underline{best}} results are highlighted.}}
\label{tab:ablation}
\resizebox{\textwidth}{!}{
\scriptsize
\begin{tabular}{cc|cc|cc|cc}
\hline
 &  & \multicolumn{6}{c}{CPU Index} \\
\cline{3-8}
Horizon & Metric 
& \multicolumn{2}{c}{Primary (US)} 
& \multicolumn{2}{c}{Alternative (US)} 
& \multicolumn{2}{c}{Global} \\
 &  & BSTS-X$_\text{MG}^\text{TI}$ & BSTS-X$_\text{MG}^\text{TV}$ & BSTS-X$_\text{MG}^\text{TI}$ & BSTS-X$_\text{MG}^\text{TV}$ & BSTS-X$_\text{MG}^\text{TI}$ & BSTS-X$_\text{MG}^\text{TV}$ \\
\hline
3  & MAPE  & \textbf{\underline{4.708}} & 26.641 & \textbf{\underline{23.856}} & 46.761 & \textbf{\underline{18.107}} & 30.275 \\
   & SMAPE & \textbf{\underline{4.598}} & 24.641  & \textbf{\underline{22.791}}  & 32.492  & \textbf{\underline{16.070}} & 26.014 \\
   & MAE   & \textbf{\underline{10.568}} & 59.926 & \textbf{\underline{0.428}}  & 0.727  & \textbf{\underline{26.115}} & 45.461 \\
   & MASE  & \textbf{\underline{1.817}} & 10.303 & \textbf{\underline{0.854}}  & 1.450  & \textbf{\underline{1.0102}} & 1.759 \\
   & RMSE  & \textbf{\underline{10.655}} & 64.737 & \textbf{\underline{0.435}}  & 0.943  & \textbf{\underline{30.484}} & 46.470 \\
\hline
6  & MAPE  & \textbf{\underline{6.029}} & 23.638 & 23.627 & \textbf{\underline{22.231}} & \textbf{\underline{16.176}} & 18.171 \\
   & SMAPE & \textbf{\underline{6.343}} & 27.664  & 23.954  & \textbf{\underline{22.034}}  & \textbf{\underline{16.179}} & 20.728 \\
   & MAE   & \textbf{\underline{15.572}} & 57.833 & 0.469  & \textbf{\underline{0.411}}  & \textbf{\underline{28.673}} & 34.636 \\
   & MASE  & \textbf{\underline{0.530}} & 1.968  & 0.890  & \textbf{\underline{0.779}}  & \textbf{\underline{1.024}} & 1.237 \\
   & RMSE  & \textbf{\underline{24.201}} & 68.115 & 0.506  & \textbf{\underline{0.483}}  & \textbf{\underline{32.601}} & 42.796 \\
\hline
12 & MAPE  & \textbf{\underline{11.345}} & 32.280 & \textbf{\underline{43.469}} & 79.300 & 15.553 & \textbf{\underline{12.755}} \\
   & SMAPE & \textbf{\underline{12.133}} & 42.321  & \textbf{\underline{44.755}}  & 113.716  & 14.596 & \textbf{\underline{13.383}} \\
   & MAE   & \textbf{\underline{29.361}} & 79.859 & \textbf{\underline{0.928}}  & 1.849  & 25.012 & \textbf{\underline{22.110}} \\
   & MASE  & \textbf{\underline{0.842}} & 2.289  & \textbf{\underline{1.224}}  & 2.438  & 1.181 & \textbf{\underline{1.044}} \\
   & RMSE  & \textbf{\underline{44.153}} & 100.536 & \textbf{\underline{1.140}} & 2.345 & 28.018 & \textbf{\underline{27.415}} \\
\hline
24 & MAPE  & \textbf{\underline{18.194}} & 33.267 & \textbf{\underline{33.092}} & 110.307 & \textbf{\underline{18.247}} & 29.277 \\
   & SMAPE & \textbf{\underline{19.182}} & 39.950  & \textbf{\underline{36.085}}  & 155.571  & \textbf{\underline{20.305}} & 36.051 \\
   & MAE   & \textbf{\underline{44.670}} & 79.877 & \textbf{\underline{1.028}}  & 3.000  & \textbf{\underline{35.001}} & 55.830 \\
   & MASE  & \textbf{\underline{0.728}} & 1.301  & \textbf{\underline{0.844}}  & 2.464  & \textbf{\underline{0.910}} & 1.452 \\
   & RMSE  & \textbf{\underline{66.440}} & 98.975 & \textbf{\underline{1.416}}  & 3.381  & \textbf{\underline{44.656}} & 66.796 \\
\hline
\end{tabular}
}
\end{table}

\subsection{Impulse Response Analysis}\label{sub:impluse_analysis}
\textcolor{black}{To empirically validate the theoretical links discussed in Sections~\ref{sec:lit_review} and~\ref{sub:motivation} and further enhance the economic intuition, this section examines the responses of the primary US CPU index to innovations in key macroeconomic and financial cycle variables. These innovations correspond to one-standard-deviation forecast errors from the estimated model rather than identified structural shocks.} The impulse response analysis is conducted using the local projections method of \cite{jorda2005estimation}, which is well-suited in this context due to its resilience to model misspecification and its flexibility in estimating dynamic effects without imposing restrictive assumptions often associated with vector autoregression (VAR) models. \textcolor{black}{Given the structural uncertainty and high dimensionality associated with the determinants of the CPU index, the local projection approach offers a transparent framework for quantifying how unexpected movements, such as an increase in unemployment or shifts in credit markets, impact CPU across various time horizons.} This approach is particularly beneficial in climate policy-related research, where grasping the timing, volatility, and persistence of policy uncertainty is essential for crafting adaptive and forward-thinking strategies. Therefore, this analysis provides deeper insights into the dynamic transmission processes between macroeconomic and financial cycle factors and the CPU index.

\sloppy{The impulse response analysis focuses on sixteen variables: business confidence index (BCI), composite leading indicator (CLI), real S\&P500 index growth (SP500), cyclically adjusted price-to-earnings ratio (CAP/Earnings), real credit growth (CRED), credit-to-GDP ratio (CRED/GDP), house price index (HPI), cyclically adjusted price-to-rent ratio (CAP/Rent), household real mortgage debt growth (Mortgage), household mortgage-to-income ratio (Mortgage/Income), private residential fixed investment-to-GDP ratio (PRFI/GDP), personal interest payments-to-income ratio (PIP/Income), unemployment rate (Unemployment Rate), total housing starts (HOUST), real personal consumption expenditures (DPCERA3M086SBEA), and new private housing permits in the northeastern US (PERMITNE). These 
variables represent the widest spectrum of the overall aggregate macroeconomy, covering major sectors, such as labor, credit, housing, consumption, and financial markets, to provide exploratory insights into how broader macroeconomic conditions influence the CPU index.}

\textcolor{black}{This analysis employs the local projections methodology on the training sample of monthly data spanning April 1987 through June 2021 to examine the dynamic impact of financial and macroeconomic innovations on the CPU index.} The estimation of the impulse response functions follows the local projections methodology originally developed by \cite{jorda2005estimation}, and computationally implemented by the \texttt{lpirfs} package in \textbf{R} \cite{adammer2019lpirfs}. This approach fundamentally differs from the VAR method by employing direct sequential regressions on future realizations of the dependent variable, thereby preventing the necessity to specify and invert a complete multivariate dynamic system \cite{jorda2005estimation}. 
This study employs linear local projections estimated via the \texttt{lp-lin} function with the following specifications: a maximum lag length of 12 months determined by the Bayesian Information Criterion (\texttt{max-lags = 12}), a linear time trend (\texttt{trend = 1}), innovations standardized to one standard deviation (\texttt{shock-type = 0}), and 95\% confidence intervals computed using heteroskedasticity and autocorrelation-robust Newey-West \cite{newey1987hypothesis} standard errors (\texttt{confint = 1.96}, \texttt{use-nw = TRUE}). The analysis estimates impulse response functions across a 24-month-ahead horizon (\texttt{hor = 24}). In this study, the CPU index measures the degree of uncertainty in climate-related policy, with higher values indicating greater uncertainty. This is essential for interpreting the impulse response results. A positive impulse response signifies an increase in CPU, while a negative response represents a reduction in uncertainty.

\begin{figure}[h!]
 \centering
  \includegraphics[width=0.9\textwidth]{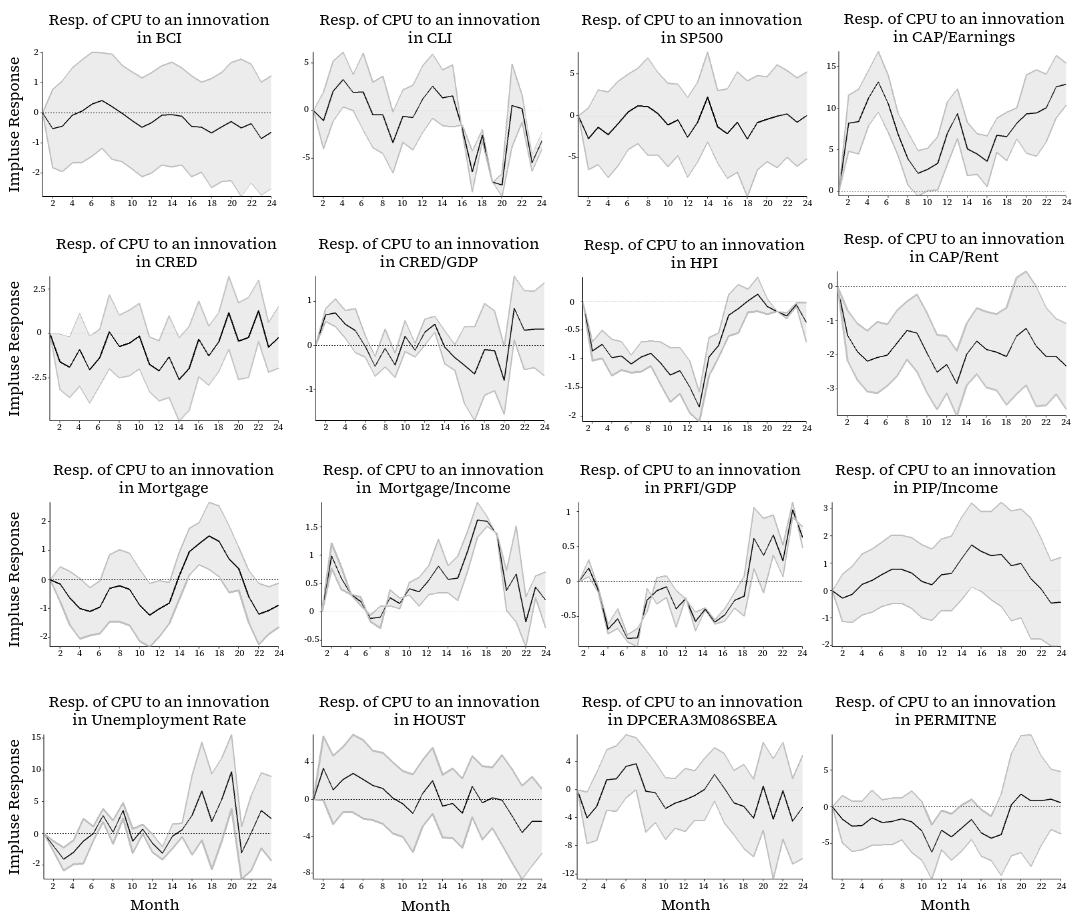}
   \caption{\textcolor{black}{Impulse response functions generated via the local projections method, quantifying the CPU index's dynamic reactions to one-standard-deviation innovations in macroeconomic and financial cycle variables. The reaction is represented by the solid black line, and the 95\% confidence intervals are represented by the gray shaded regions. The dashed black line displays the zero line. The sample period runs from April 1987 to June 2021.}} 
   \label{fig:IRF}
\end{figure}

Fig.~\ref{fig:IRF} illustrates the dynamic responses of the CPU index to innovations in macroeconomic and financial cycle variables across a 24-month-ahead horizon. The impulse response analysis shows that the variables have economically meaningful and statistically significant effects on the CPU index, either positively or negatively. Among the variables that have positive influences, CAP/Earnings exhibits a consistently positive effect over the 24-month-ahead horizon, albeit with some fluctuations, indicating a persistent and substantial relationship in which equity market overvaluation increases CPU. A positive relation between CAP/Earnings and CPU suggests that when the markets have high valuations, any regulation change, including climate-related policies such as an increase in environmental taxes or upgradation of production facilities, can increase the companies' costs and reduce their profitability. Reduced earnings in the period of overvalued financial markets lead to a sharp market correction. Hence, during market overvaluation, the investors can be very sensitive to changes in climate-related regulatory policies, thereby increasing the perceived uncertainty. The impulse response of CPU to PIP/Income is also positive. This positive relationship implies that when households face higher interest payments relative to their income to service debt, they become more sensitive to regulatory changes, including those related to climate policy. This reflects a tension between the increased financial vulnerability of individual households and the perceived cost of implementing climate policy regulations. \textcolor{black}{Similarly, Mortgage/Income exhibits a predominantly positive effect over the 24-month-ahead horizon, indicating that higher household debt burdens amplify financial vulnerability and public sensitivity to climate-related regulatory changes.} In addition, CRED/GDP shows a predominantly positive effect, indicating a sustained relationship between systemic leverage and CPU. An elevated level of credit relative to GDP heightens financial vulnerability, which may discourage policymakers from implementing substantial policy changes that could destabilize corporate financial situations \cite{annicchiarico2022business}. Adoption and compliance with new climate regulations typically require significant business investment. In an environment of excessive credit growth relative to GDP, central banks may implement stricter macroprudential policies to contain systemic risk. Such tightening can limit firms’ capacity to invest in compliance with climate regulations, thereby heightening CPU.

Further supporting the sensitivity of CPU to cyclical macroeconomic conditions is its mostly positive response to the unemployment rate. When the unemployment rate increases, the focus shifts from environmental concerns to basic survival needs. This again makes households reluctant to comply with any new environmental policies, thereby making the introduction and implementation of any changes to climate policy regulation very difficult and politically sensitive. Lastly, we observe a positive response with PRFI/GDP, though the effect occurs in the first three months and reemerges during the final six months of the 24-month horizon. A greater PRFI/GDP indicates a rise in residential construction, which may lead to increased environmental consequences such as elevated energy consumption, land development, and construction-related emissions. This may drive policymakers to propose stricter climate regulations to alleviate the threats to environmental or climatic degradation. The polices may include stricter emission controls, increased taxes, or changes in the building standards. This expectation of any forthcoming rigorous and stringent policies can increase uncertainty surrounding climate policy, regarding their timing and intensity, thereby positively influencing the CPU index \cite{cho2024climate}. 

Now we shift our focus to the variables that exert a negative and statistically significant influence on the CPU index. CAP/Rent depicts an intensely negative and consistent impact throughout the 24-month-ahead horizon. The economic intuition behind this effect's sustained nature is that an increase in property prices relative to rental income reflects an increase in the net worth of the homeowners and creates a wealth effect. An increase in the real estate asset price can be linked with reduced uncertainty and increased investor confidence, making them more optimistic about the future economic outlook. Increased optimism about the future economic outlook, coupled with increased wealth, makes individuals and businesses more adaptive to climate change policies. Similarly, HPI demonstrates a negative and statistically significant response. The negative response suggests that increasing house prices signal economic stability and rising housing demand. This improved economic stability may reduce CPU as people are expected to have a higher disposable income to comply with new environmental or climate regulations. Hence, policymakers experience both lower resistance and political pressure to alter existing environmental policies \cite{obani2016impact,bumann2021determinants}. \textcolor{black}{Moreover, Mortgage shows a predominantly negative impact on CPU over the 24-month-ahead horizon, as stable financing conditions reduce household financial stress and dampen sensitivity to climate-related regulatory changes. In addition, PERMITNE exhibits a largely negative effect, suggesting that stronger forward-looking construction activity signals macroeconomic resilience and greater household and firm adaptability to climate regulations. Similarly, HOUST shows a negative response despite an initially positive effect, indicating that while short-run increases in construction activity may raise regulatory concerns, sustained housing expansion ultimately reduces uncertainty as income and investment conditions improve \cite{cho2024climate}.}

The negative response of the CPU index to CRED can be attributed to the fact that business cycles alternative between economic recovery and recession. When the economy enters the recovery phase following a recession or stagnation, an initial wave of credit growth typically emerges, driven by rising investment and employment opportunities at the start of the new cycle. This period is often marked by robust corporate profits and growing household incomes, thereby creating a favorable environment for policymakers to simultaneously introduce and implement ambitious climate-related policies. However, when credit growth exceeds the growth rate of economic output, the resulting increase in credit can heighten climate-related uncertainties, as previously discussed in the case of CRED/GDP. \textcolor{black}{The business confidence measures, BCI and CLI, exhibit predominantly negative responses over the 24-month-ahead horizon, indicating that stronger forward-looking economic expectations can reduce CPU, although its response to CLI fluctuates, reflecting its sensitivity to cyclical conditions \cite{zhang2020country}. Similarly, DPCERA3M086SBEA exhibits a mostly negative effect, suggesting that stronger consumption reflects economic expansion and public confidence, thereby creating political space for the implementation of climate policies with lower perceived uncertainty \cite{zhang2022influence}.} Regarding the S\&P 500 index, it demonstrates a negative and statistically significant impact on CPU. S\&P 500 acts as a sentiment barometer for the aggregate macroeconomy, which influences the perceived viability of climate-friendly and environmental investments. Rising stock prices in equity markets lead to increased wealth in the hands of investors. The positive wealth effect creates positive sentiment about the economy among investors. This forward-looking optimism enhances the commitment of households and businesses towards adherence to any new changes in environmental and climate change regulations. Such enhanced commitment and positive economic outlook give an opportunity to policymakers to pursue and implement new climate change policies. The opposite scenario occurs during a recession-driven bear market when negative forward-looking behaviour makes it difficult for policymakers to implement stricter climate policies \cite{khan2019carbon, amin2021role}.

\textcolor{black}{The impulse response analysis results indicate that the effects of macro-financial conditions on CPU are not only statistically significant but also economically meaningful and persistent over time. Periods characterized by financial stability, rising asset prices, and strong confidence provide governments with greater political and fiscal capacity to sustain long-term climate commitments. Favorable economic conditions create a buffer that facilitates credible and consistent policy communication, as households and firms are better positioned to absorb the adjustment costs associated with the green transition. By contrast, during episodes of financial stress or economic contraction, heightened financial vulnerability and weakened confidence constrain policymakers’ scope for action, shifting priorities toward short-term stabilization. Consequently, environmental commitments are more likely to be delayed, diluted, or reframed as governments adopt a reactive stance to mitigate immediate economic pressures. This cyclical reorientation of policy objectives leads to inconsistencies in climate policy direction, which manifests as pronounced increases in CPU.} These patterns underscore that CPU is closely linked to the underlying condition of the state of the economy.

\subsection{Economic Interpretation of Feature Importance Plot}\label{feature_plot}
\begin{figure}[h!]
 \centering
  \includegraphics[width=0.9\textwidth]{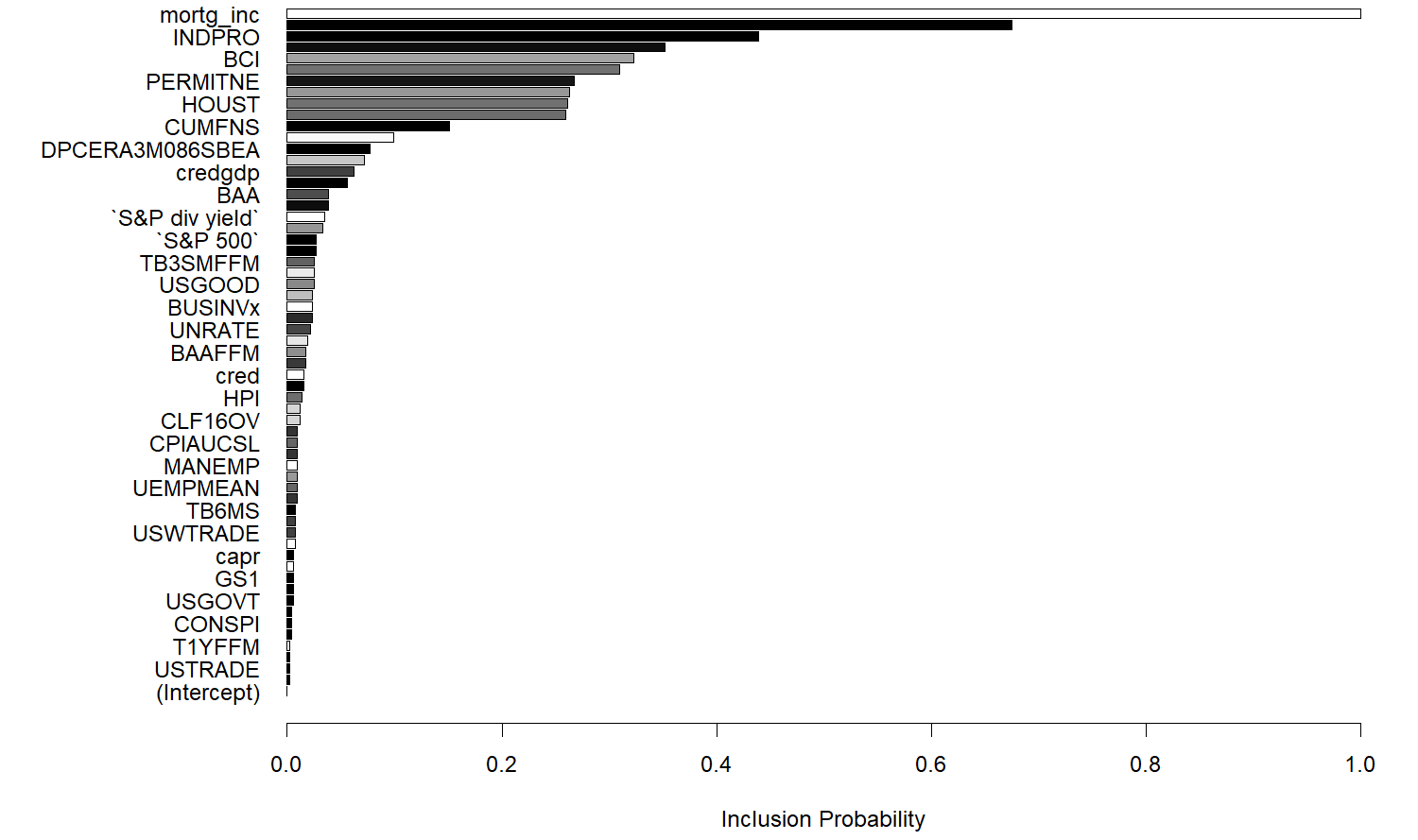}
   \caption{Inclusion probabilities of variables selected by the BSTS model over a 24-month-ahead forecast horizon. The plot reflects the relative importance of each variable in contributing to the model’s predictive accuracy, with higher probabilities indicating greater explanatory relevance. The full form of the variables displayed on the y-axis is detailed in Table~\ref{tab:causal_analyses}} 
   \label{fig:BSTS-Variables}
\end{figure}

In Sections~\ref{results} and~\ref{stat_sig}, we highlighted the superior performance of the BSTS model compared to both classical and modern forecasting architectures. In this section, we provide insight into the superior performance of the BSTS model by examining its variable selection mechanism. For interpretability, we focus on the macroeconomic and financial cycle variables, excluding the Google Trends indicators. Nonetheless, their relevance remains clear from the preceding results and Fig.~\ref{fig:google}, where the time series of several climate-related search terms closely mirror the evolution of CPU, underscoring their complementary role in capturing public sentiment surrounding CPU. To examine whether the variables selected by the BSTS model align with economic theory in Section~\ref{sub:motivation} and empirical impulse response analysis in Section~\ref{sub:impluse_analysis}, we analyze the feature importance plot generated by the BSTS model over the 24-month forecast horizon. Fig.~\ref{fig:BSTS-Variables} highlights the variables with the highest inclusion probabilities, reflecting broad sectors of the economy, such as financial markets, housing, labor, and economic sentiment, that are theoretically linked to CPU.

The top ten variables selected by the model, as illustrated in Fig.~\ref{fig:BSTS-Variables}, exhibit the highest inclusion probabilities. Importantly, these variables are consistent with economic theory and with the factors previously identified as significant determinants of CPU in Sections~\ref{sub:motivation} and~\ref{sub:impluse_analysis}. At the top of the list is the mortgage-to-income ratio ($\text{mortg}\_\text{inc}$), which captures household financial vulnerability to interest rate changes when central banks increase rates. Higher mortgage burdens reduce disposable income, thereby heightening public resistance to policies that might increase living costs. Such resistance, in turn, can pressure policymakers to postpone or soften environmental initiatives. The industrial production index (INDPRO) monitors real economic output across major industrial production sectors of the US and plays an important role due to its direct correlation with energy consumption and emissions. Fluctuations in industrial activity can deter firms from adopting new environmental regulations, complicating policymakers’ efforts to balance economic growth with environmental sustainability. Similarly, the business confidence index (BCI) captures firms’ expectations about future economic conditions and investment decisions, including their willingness to commit resources toward compliance with green policies \cite{zhang2020country}. Together, INDPRO and BCI represent the current and forward-looking dimensions of business sector performance and are tied to CPU.

The model also gives high inclusion probabilities to housing sector indicators, such as new private housing permits in the northeastern US (PERMITNE) and total housing starts (HOUST). Their inclusion is consistent with economic theory and earlier findings in this study, confirming the macroeconomic relevance of the housing sector in shaping the CPU index. High values of PERMITNE and HOUST reflect a robust construction sector and signal overall economic vitality. In such periods of optimism, governments face fewer constraints in implementing long-term climate policies, as households are more financially secure and more receptive to adopting climate-friendly technologies, such as energy-efficient installations in new homes. Conversely, a decline in housing permits and new home construction signals economic weakness or stagnation. Under such conditions, policymakers may delay or soften environmental regulations to avoid further straining the housing and real estate sector. As highlighted by \cite{cho2024climate}, stringent climate regulations can amplify vulnerabilities in an already fragile housing market, underscoring the delicate balance policymakers must strike between environmental and macroeconomic stability. The manufacturing capacity utilization (CUMFNS) measures the extent to which existing productive capacity is being employed. High utilization rates generate resource constraints and inflationary pressures, prompting interest rate hikes, whereas low utilization motivates monetary easing to stimulate growth. Lower interest rates, in turn, 
accelerate the transition toward clean energy since lowering financing costs 
enables firms to invest more readily in environmentally sustainable technologies. However, the inclusion probability of CUMFNS is notably lower than that of PERMITNE and HOUST, which is consistent with our earlier discussion, in Section~\ref{sub:motivation}, that housing indicators exert a stronger influence on CPU dynamics than sector-specific production measures.

The inclusion of real personal consumption expenditures (DPCERA3M086SBEA) and the credit-to-GDP ratio (credgdp) further confirms the consistency of the model’s variable selection with economic reasoning. As discussed earlier, both variables play a central role in shaping macro-financial dynamics and, consequently, CPU. Consumer spending is a key driver of economic activity in the US, reflecting household confidence and overall economic health. A sustained rise in real personal consumption signals economic prosperity, enabling governments to pursue more ambitious climate actions. Conversely, a contraction in consumption coincides with heightened economic uncertainty, which can lead to delays or moderation in climate policy initiatives \cite{zhang2022influence}. Similarly, the credit-to-GDP ratio captures the level of financial leverage within the economy. Elevated levels increase systemic vulnerability, making policymakers more cautious about introducing policies that could disrupt credit markets. During such periods, central banks tighten macro-prudential regulations to contain financial risks, which indirectly slows the pace of climate policy implementation by limiting investment and credit availability for green projects. The indicators S\&P 500 dividend yield and Moody’s Seasoned Baa Corporate Bond Yield (BAA) capture capital market conditions, credit risk, and market sentiment, all of which are crucial in shaping CPU. Increases in bond yields or high dividend yields signal investor caution and a shift toward risk-averse behavior, particularly in capital-intensive and regulated industries. Such caution can heighten market sensitivity to policy changes, leading investors and firms to delay climate-related investments or push back against stringent policy measures.

The ten factors identified by the model span broad dimensions of the economy, capturing sectoral exposure to policy risk and macro-financial stability \cite{annicchiarico2022business, giovanardi2024pro}. All of these factors have been found to have a significant impact on CPU. This coherence with economic theory provides a strong justification for the model’s superior predictive performance, as it successfully isolates the most economically relevant predictors of CPU. We also observe that several variables previously identified as theoretically relevant to CPU appear in Fig.~\ref{fig:BSTS-Variables} with smaller inclusion probabilities. These include S\&P 500 index, unemployment rate (UNRATE), credit growth (cred), real house price growth (HPI), and the cyclically adjusted price-to-rent ratio (capr). Their limited inclusion probabilities suggest that while they contain information related to CPU, their explanatory contribution is largely captured by more dominant macro-financial variables already included in the model, such as credit-to-GDP ratio, INDPRO, and the housing indicators (PERMITNE and HOUST). It is worth noting that, despite their intuitive relevance, sector-specific variables such as mining employment do not appear among the top predictors in Fig.~\ref{fig:BSTS-Variables}. This outcome is economically consistent with how CPU propagates through the economy. CPU is not confined to specific industries but reflects broader expectations, confidence, and investment behavior across households, firms, and financial markets. Thus, aggregate indicators, such as housing activity, credit conditions, and financial market sentiment, capture the nature of policy uncertainty more effectively than sectoral metrics in individual industries. 

Interestingly, other related variables, such as the 
3-month treasury bill minus federal funds rate (TB3SMFFM), all employees in goods-producing industries (USGOOD), and the consumer price index (CPI) for all urban consumers (CPIAUCSL), appear in the feature importance plot with small but non-zero inclusion probabilities. These can be interpreted as complementary measures that reflect similar underlying mechanisms to those captured by the earlier-identified but weakly included variables. For instance, treasury spreads and short-term interest rate differentials serve as proxies for financial conditions and market expectations, linking indirectly to credit growth and equity market sentiment. Likewise, variables such as manufacturing employment (MANEMP), wholesale and retail trade employment (USWTRADE and USTRADE), and the civilian labor force level (CLF16OV) capture aggregate labor market dynamics that are closely related to the unemployment rate, whereas CPI provides an alternative channel for capturing household purchasing power and cost-of-living pressures that affect public support for climate policies. In contrast, some variables we initially considered relevant, such as the composite leading indicator, private residential fixed investment-to-GDP ratio, and personal interest payments-to-income ratio, do not appear in Fig.~\ref{fig:BSTS-Variables}. This likely reflects multicollinearity, as their effects are proxied by broader housing and credit indicators that provide more stable predictive signals. Overall, this analysis confirms that the model's variable selection mechanism identifies predictors consistent with both economic theory and our empirical findings, with inclusion probabilities clearly differentiating the primary drivers of CPU from secondary contributors.

\subsection{Uncertainty Quantification using Credible Intervals}\label{credible_intervals}
Finally, we aim to quantify the uncertainty associated with the BSTS forecasts by analyzing the model's credible intervals. In Bayesian statistics, a credible interval is the range within which a forecast lies with a certain probability, given the posterior distribution. Unlike frequentist confidence intervals, which are based on hypothetical repeated sampling, credible intervals are directly derived from the posterior distribution, incorporating both prior knowledge and observed data. Let \(y_{1:T}\) represent the observed time series up to time \(T\). The predictive distribution for \(y_{T+h}\) at horizon \(h\) conditional on the past observations \(y_{1:T}\) is expressed as:
\[
p(y_{T+h} \mid y_{1:T}) = \int p(y_{T+h} \mid \theta, y_{1:T}) \, p(\theta \mid y_{1:T}) \, d\theta,
\]
where \(p(y_{T+h} \mid \theta, y_{1:T})\) is the likelihood of the future observation given model parameters \(\theta\), and \(p(\theta \mid y_{1:T})\) is the posterior distribution of the parameters after observing the data \(y_{1:T}\). This integration captures both the uncertainty in the future outcomes and the model parameters. 

The \(100(1-\alpha)\%\) credible interval for the forecast at \(T+h\) is defined as the interval within which \(y_{T+h}\) lies with posterior probability \(1-\alpha\), such that \(
P(L_{T+h} \leq y_{T+h} \leq U_{T+h} \mid y_{1:T}) = 1 - \alpha\), where \(L_{T+h}\) and \(U_{T+h}\) denote the lower and upper bounds of the credible interval, respectively. These bounds correspond to the quantiles of the posterior predictive distribution, expressed as \(L_{T+h} = F^{-1}(\alpha/2)\) and \(U_{T+h} = F^{-1}(1-\alpha/2)\), with \(F^{-1}(\cdot)\) denoting the inverse cumulative distribution function of the predictive distribution. In practice, since the posterior distribution is usually approximated via Monte Carlo sampling, the credible interval is obtained from the empirical quantiles of the simulated forecasts as \(L_{T+h} = \hat{y}_{T+h}^{(m_{\alpha/2})}\) and \(U_{T+h} = \hat{y}_{T+h}^{(m_{1-\alpha/2})}\), where \(\hat{y}_{T+h}^{(m)}\) denotes the forecast from the \(m\)-th draw, and \(m_{\alpha/2}\) and \(m_{1-\alpha/2}\) correspond to the empirical quantiles of the posterior sample at the \(\alpha/2\) and \(1-\alpha/2\) levels, respectively \cite{ohagan1994bayesian}.

\begin{figure}[h!]
 \centering
 \includegraphics[width=0.9\textwidth]{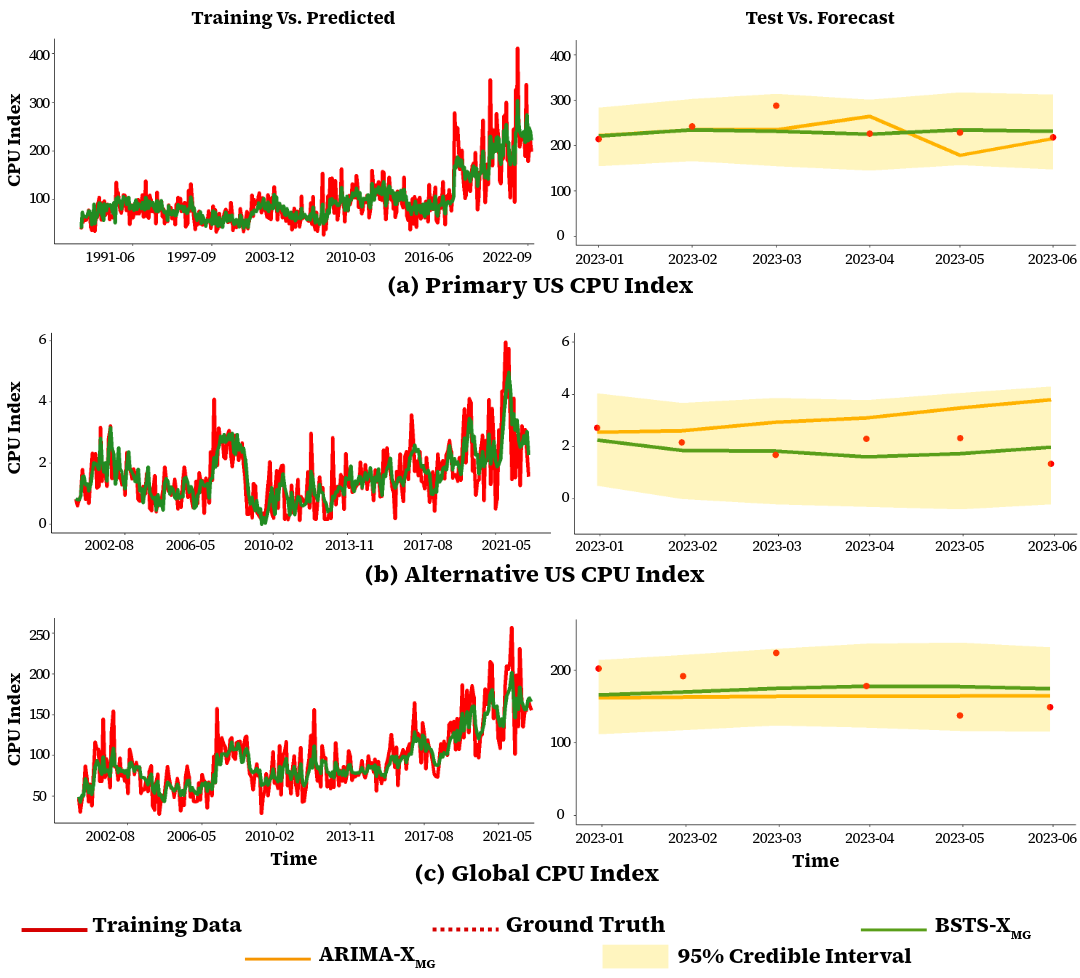}
 \caption{\textcolor{black}{Visualization of the ground truth (red), fitted values and forecasts of the BSTS-X$_\text{MG}$ model (green), forecasts of ARIMA-X$_\text{MG}$ (orange), and the credible intervals generated by BSTS (yellow shaded) for the 6-month holdout period of the (a) primary US, (b) alternative US, and (c) Global CPU series. The plot on the left displays the training data along with the fitted values of BSTS-X$_\text{MG}$. The plot on the right presents the semi-long-term forecasts shown as both point estimates and credible intervals.}}
   \label{fig:CPI-6}
\end{figure}

Fig.~\ref{fig:CPI-6} presents the 6-month-ahead forecasts for the CPU index across the three series produced by the BSTS-X$_\text{MG}$ model. The left panel displays the training series alongside the fitted values, illustrating the model’s strong in-sample fit and its ability to replicate the historical dynamics of the CPU index. \textcolor{black}{The right panel depicts the point forecasts and the 95\% credible intervals of the BSTS-X$_\text{MG}$ together with the point forecasts from ARIMA-X$_\text{MG}$. Although ARIMA-X$_\text{MG}$ demonstrates strong performance, it fails to adequately capture the volatility observed in the CPU index during the test period. This limited responsiveness to evolving patterns underscores its weakness for policy uncertainty forecasting.} In contrast, BSTS-X$_\text{MG}$ achieves closer alignment with the ground truth and demonstrates superior adaptability to the series’ dynamic behavior. This adaptability is further strengthened by the model’s explicit quantification of forecast uncertainty through credible intervals derived directly from the posterior predictive distribution. 
These intervals provide a probabilistic range within which future CPU values are expected to lie, reflecting the model’s ability to realistically account for substantial uncertainty in climate-related policy dynamics. The credible intervals enable policymakers to better understand the range of plausible future outcomes and to plan accordingly under uncertainty, thus making BSTS a more informative and robust forecasting framework. Overall, the capacity of BSTS to adapt to changing trends while rigorously quantifying uncertainty establishes it as a reliable tool for forecasting policy-sensitive and volatile series such as the CPU index.

\section{Policy Implication}\label{sec:discussion}
\textcolor{black}{This study uncovers a complex relationship between macro-financial conditions and the dynamics of the CPU index, with clear differences in the relative influence of key predictors. The most influential factors are household financial vulnerability (mortgage-to-income ratio), business confidence (BCI), and housing activity (PERMITNE, HOUST).} High mortgage burdens signal financial strain, which can increase public resistance to policies perceived as raising living costs, thereby amplifying CPU. Forward-looking indicators such as BCI capture expected business and household sentiment; elevated confidence tends to reduce CPU by fostering investment and regulatory compliance, while low confidence signals economic stress, leading governments to prioritize short-term recovery over long-term climate goals. Housing indicators, including PERMITNE and HOUST, reflect broader economic vitality; strong activity provides policymakers more latitude to implement stringent climate measures, whereas downturns constrain policy action \cite{cho2024climate}.

\textcolor{black}{Secondary predictors provide additional, though less dominant, information. These include industrial production, manufacturing capacity utilization, the real personal consumption, credit-to-GDP ratio, and market valuations such as S\&P 500 dividend yields and Moody’s Baa bond yields.} For example, higher asset prices or home price indices are generally associated with lower CPU, as stronger asset performance fosters optimism and reduces perceived policy risk. Conversely, elevated financial leverage, overvalued markets, or constrained credit conditions increase uncertainty by heightening sensitivity to regulatory changes because such monetary tightening can deter firms and households from investing in green technologies, thereby increasing uncertainty about future climate policy directions. \textcolor{black}{Other variables, including short-term treasury spreads, labor market indicators, CPI, and sector-specific employment measures, provide complementary information but exert smaller effects. Their lower inclusion probabilities, as depicted in Fig~\ref{fig:BSTS-Variables}, indicate that their contributions are largely captured by dominant macro-financial variables such as housing, credit, and confidence indicators.}

\textcolor{black}{Dynamic local projection analysis further quantifies these associations over time. Higher cyclically adjusted price-to-earnings, household debt burdens (mortgage-to-income, personal interest payments-to-income), and credit-to-GDP ratios are associated with increases in CPU, reflecting greater sensitivity to regulatory changes during periods of financial vulnerability or overvalued markets. In contrast, stronger housing activity (PERMITNE, HOUST), rising property prices (HPI, cyclically adjusted price-to-rent ratio), elevated business confidence (BCI, CLI), personal consumption growth, and equity prices are associated with reductions in CPU, indicating that healthier macro-financial conditions facilitate policy implementation.} These patterns confirm that CPU intensifies during downturns and moderates during expansions, consistent with prior evidence linking recessions to weaker environmental policy commitment \cite{ide2020recession}. Public sentiment, captured through Google Trends indicators, also plays an equally important role. Search intensity for climate-related terms provides a quantifiable proxy for societal attention and concern. These sentiment-based indicators reflect early shifts in collective awareness that often precede policy debates or legislative actions. Their inclusion significantly improved long-term forecast accuracy, underscoring the value of integrating behavioral dimensions into traditional economic forecasting frameworks.

\textcolor{black}{Our results indicate that CPU is not only a phenomenon dependent on macroeconomic cycles but also a governance challenge at the implementation level. The strong predictive power of macroeconomic variables suggests that rigid regulatory interventions may be ineffective during recessions. Instead, climate policies should be adaptive, incorporating phased compliance, conditional enforcement triggers, and pre-specified review clauses. These measures should preserve long-term climate commitments, while allowing short-term flexibility in execution. Such features will enhance policy credibility and reduce the risk of abrupt reversals. This approach directly addresses uncertainty faced by regulatory bodies, firms, and individuals, and mitigates inconsistencies in policy enforcement during periods of economic stress. }

\textcolor{black}{Our findings further demonstrate that effective implementation of environmental policies requires coordination from entities beyond environmental agencies alone. The significant influence of housing, credit, and household financial stress variables on CPU suggests that climate policy implementation depends on broader economic governance structures. Consequently, climate regulations will be more effective when they are coordinated with financial regulation, housing policy, and fiscal support mechanisms. For example, authorities should design interest rate-sensitive policy instruments that give more robust subsidies or low-interest financing for green technologies when market credit conditions tighten, thereby maintaining steady progress even during tight liquidity conditions. The adaptability of households and firms to environmental mandates can be further increased if they are integrated with green finance instruments, targeted subsidies, or credit guarantees. In economic downturns, instead of rolling back climate regulations,  green fiscal stimuli should be provided. These green stimuli will motivate firms to support continued investment in environmental technologies, preventing a ``lost decade" for climate action. Such integration of climate policies with macroeconomic frameworks can substantially lower the adjustment costs.}

\textcolor{black}{This study underscores the importance of expectation management and information governance in mitigating CPU. The significant role of sentiment-based indicators implies that uncertainty reflects ambiguity about policy direction and its credibility rather than the absence of policy. Hence, it is imperative for environmental authorities to address policy uncertainty with transparent communication strategies, which include providing forward guidance on regulatory pathways, predictable rulemaking schedules, and clear signaling of long-term policy continuity. Such measures are indispensable for strengthening institutional trust among the stakeholders and reducing informational frictions, which is critical for effective policy implementation in politically and economically volatile environments.}

\textcolor{black}{Finally, the probabilistic forecasting framework employed in this study provides a practical tool for risk-aware environmental planning. Rather than relying solely on deterministic projections, environmental managers can leverage forecast distributions of CPU to identify periods of heightened implementation risk. Accordingly, they can design contingency measures to enhance the implementation of climate-related regulations. Integrating uncertainty quantification into policy evaluation enables regulators to anticipate challenges in implementing climate policies, prioritize interventions, and sustain climate objectives across economic cycles. This integrated approach shifts environmental management from a reactive to a forward-looking, robust governance framework that explicitly accounts for economic and political uncertainty.}


\section{Conclusion}\label{sec:conclusion}
\textcolor{black}{This study provides the first systematic attempt to forecast US and global CPU, shifting attention from its consequences to the identification of its underlying predictors. To this end, we integrate four empirical screening techniques to rigorously identify macro-financial variables most strongly associated with CPU. The selected predictors are not only statistically significant but also theoretically grounded, aligning with economic reasoning regarding household financial vulnerability, business confidence, and housing activity, among other channels. To quantify the dynamic influence of these predictors on CPU, we employ local projection impulse response analysis. This approach reveals that increases in asset prices, business confidence, and consumption reduce CPU, whereas elevated financial leverage, high debt burdens, or overvalued markets amplify uncertainty. These results highlight the cyclical nature of CPU, which intensifies during downturns and moderates during expansions, and demonstrate how macro-financial conditions propagate through the economy to influence CPU over time.}

\textcolor{black}{Recognizing that CPU also reflects public perception, we incorporate sentiment-based indicators derived from Google Trends. These measures capture early shifts in societal attention to climate policies and empirically improve forecast accuracy, underscoring the value of integrating behavioral dimensions alongside traditional economic indicators. Using these variables, we forecast both US and Global CPU indices, providing a comprehensive assessment of predictive performance and model robustness across indices. Our extensive experiments demonstrate that BSTS with time-invariant regression coefficients achieves the highest forecasting accuracy for all CPU indices. Importantly, this analysis demonstrates that BSTS's forecasting accuracy is not coincidental but stems from its principled selection of variables whose impacts are consistent with economic theory and validated empirically through dynamic response analysis. This link between theory and predictive power is illustrated in the feature importance plot, which identifies household financial vulnerability (mortgage-to-income ratio), business confidence (BCI), and housing activity (PERMITNE, HOUST) as the variables with the largest dynamic impacts on CPU, while secondary predictors, such as industrial production, credit-to-GDP ratios, and market valuations, provide complementary information that captures how uncertainty propagates through macro-financial channels.}

\textcolor{black}{From a policy perspective, these findings indicate that rigid interventions may be less effective during economic downturns. Climate policies should therefore be adaptive, incorporating phased compliance, conditional enforcement, and pre-specified review mechanisms to balance long-term objectives with short-term flexibility. Implementation is most effective when coordinated across housing, credit, and fiscal policy, and supported by green financing instruments and targeted fiscal measures to sustain progress under economic stress. Clear communication, forward guidance, and predictable rulemaking enhance institutional trust and reduce informational frictions, helping stabilize CPU. Finally, probabilistic forecasts of CPU allow regulators to anticipate high-risk periods, prioritize interventions, and maintain climate objectives consistently across economic cycles. Overall, this study demonstrates that accurate CPU forecasting requires a careful combination of rigorous variable selection, theoretical validation, and dynamic quantification of predictor effects. By uncovering the key predictors of CPU and integrating sentiment-based indicators, we provide the first empirically validated, theoretically grounded, and robust approach to anticipating CPU. These findings underscore the importance of aligning forecasting methodologies with economic theory, showing that predictive success is inseparable from understanding the underlying mechanisms driving uncertainty.}

One limitation of this study concerns the geographic scope of the analysis. Although both US and Global CPU indices are analyzed, the findings remain largely US-centric, which may limit their applicability to regions with different macroeconomic structures, financial systems, or climate policy environments. Expanding the study to incorporate additional regional or country-level indicators could further validate the robustness and applicability of the results and provide a more comprehensive understanding of the predictors of CPU worldwide. Another limitation lies in not incorporating political or environmental factors that may also play a role in shaping CPU. For example, political uncertainty, such as election cycles or government transitions, may influence CPU, as it often coincides with fluctuations in consumer confidence and broader economic uncertainty. Similarly, environmental factors, including extreme weather events, pollution, wildfires, and carbon emission metrics, could exert substantial influence on policy uncertainty dynamics. Future research could incorporate such factors to allow for a more comprehensive assessment of the relative contributions of macro-financial, political, and environmental predictors of CPU.

\section*{Code and Data Availability}
The primary US CPU index used in this study was obtained from the Economic Policy Uncertainty website: \url{https://www.policyuncertainty.com/climate_uncertainty.html}. \textcolor{black}{The alternative US and Global CPU indices used in this study were sourced from the China Energy and Environmental Policy Research Center Network website: \url{http://www.cnefn.com/data/download/climate-risk-database}.} For transparency, all datasets, including the statistically significant macro-financial predictors and Google Trends indicators, as well as the code necessary to reproduce the findings presented in this paper are made available on Github: \url{https://github.com/Donia-212/Climate_Policy_Uncertainty_Forecasting}.

\section*{Acknowledgement}
The authors would like to acknowledge the Associate Editor and two experienced reviewers for insightful suggestions that improved the paper. 
The authors want to thank Madhurima Panja of Sorbonne University Abu Dhabi for the discussion and help with the graphics presented in this manuscript. 



\bibliographystyle{plain}
\biboptions{square}
\bibliography{Bibliography}



\appendix
\renewcommand{\thesection}{\Alph{section}}

\section{Variable Screening and Association Analysis}\label{Causal_anal}
This section examines the relationships between both macroeconomic and financial cycle variables, as well as public attention indicators derived from Google Trends and the primary US CPU index. \textcolor{black}{To identify the most relevant predictors, we apply four complementary screening methods: Transfer Entropy, Granger Causality, Cross-Correlation, and Wavelet Coherence. These techniques capture nonlinear and linear dependencies as well as time-varying and frequency-specific associations between CPU and the candidate explanatory variables.} We briefly introduce each of these screening techniques in Sections~\ref{sec:te}-\ref{sec:wc}, and the results are reported in Section~\ref{causal_ana_res}. \textcolor{black}{The findings are interpreted as predictive associations rather than causal effects.}

\subsection{Transfer Entropy}\label{sec:te}
Information entropy was first introduced by \cite{shannon1948mathematical}. This concept is central to measuring the amount of uncertainty or information contained in a system. Within the framework of Shannon's theory, for a coupled system \((X, Y)\), where \(P_Y(y)\) is the probability density function (pdf) of the random variable \(Y\), and \(P_{X,Y}\) is the joint pdf of \(X\) and \(Y\), the joint entropy between \(X\) and \(Y\) is defined as:
\[
H(X, Y) = - \sum_{x \in X} \sum_{y \in Y} P_{X,Y}(x, y) \: \log (P_{X,Y}(x, y)).
\] 
The conditional entropy is given by $H(Y \mid X) = H(X, Y) - H(X)$, and can be interpreted as the uncertainty in \(Y\) given the knowledge of a specific value of \(X\). Transfer Entropy, introduced by \cite{schreiber2000measuring}, has been proven to be an effective tool in identifying causal relationships in nonlinear systems. It captures the directional flow of information between systems without the need for a predefined functional form of interaction. Transfer Entropy is defined as the difference between two conditional entropies:
\[
TE(X \rightarrow Y \mid Z) = H(Y^F \mid Y^P, Z^P) - H(Y^F \mid X^P, Y^P, Z^P),
\]
where \(Y^F\) is the forward time-shifted version of \(Y\) at lag $\Delta t$ with respect to past values of \(X^P\), \(Y^P\), and \(Z^P\). This may also be expressed as a sum of Shannon entropies:
\[
TE(X \rightarrow Y) = H(Y^P, X^P) - H(Y^F, Y^P, X^P) + H(Y^F, Y^P) - H(Y^P).
\]
Since Transfer Entropy is an asymmetric measure, that is, \(TE(X \rightarrow Y) \neq TE(Y \rightarrow X)\)), it can be used to quantify the direction of information flow between systems. The net information flow is defined as:
\[
\hat{TE}_{X \rightarrow Y} = TE_{X \rightarrow Y} - TE_{Y \rightarrow X}.
\]
This quantity indicates the dominant direction of information flow, with a positive value signifying that \(X\) provides more predictive information about \(Y\) than \(Y\) does about \(X\) \cite{michalowicz2013handbook}.

\subsection{Granger Causality}\label{sec:gc}
Granger causality test is widely used for analyzing predictive power between different time series. \cite{granger1969investigating} defined Granger causality as the ability to predict future values of the variable \(Y\) using past values of both \(X\) and \(Y\). In this framework, \(X\) is said to Granger-cause \(Y\) if the inclusion of past values of \(X\) improves the prediction of \(Y\). Let \(X_t\) and \(Y_t\) be random variables at time \(t\), and let \(X_t, Y_t, Z_t\) represent three stochastic processes. Define \(\hat{Y}_{t+1}\) as the predictor of \(Y\) at time \(t+1\). We evaluate the expected value of a loss function \(g(e)\), where the error is \(e = \hat{Y}_{t+1} - Y_{t+1}\), for both models. Typically, the forms of \(g\) include the \(L_1\) or \(L_2\) norms, and the functions \(f_1\) and \(f_2\) minimize the expected value of the loss function:
\[
R(Y_{t+1} \mid Y_t, Z_t) = \mathbb{E}[g(Y_{t+1} - f_1(Y_t, Z_t))]
\]
\[
R(Y_{t+1} \mid X_t, Y_t, Z_t) = \mathbb{E}[g(Y_{t+1} - f_2(X_t, Y_t, Z_t))].
\]
\begin{definition}
\label{def:granger-causality} \(X\) does not G-cause \(Y\) relative to the additional information \(Z\) if and only if:
\[
R(Y_{t+1} | X_t, Y_t, Z_t) = R(Y_{t+1} | Y_t, Z_t).
\]
\end{definition}
The classical Granger test is implemented via vector autoregressive (VAR$(p)$) model, where \(p\) is the number of lagged observations:
\[
Y(t) = \alpha + \sum_{\Delta t=1}^{p} \beta_{\Delta t} Y(t - \Delta t) + \epsilon_t,
\]
\[
Y(t) = \hat{\alpha} + \sum_{\Delta t=1}^{p} \hat{\beta}_{\Delta t} Y(t - \Delta t) + \sum_{\Delta t=1}^{p} \hat{\gamma}_{\Delta t} X(t - \Delta t) + \hat{\epsilon}_t,
\]
By Definition~\ref{def:granger-causality}, \(X\) does not G-cause \(Y\) if and only if the prediction errors from the restricted and unrestricted models are equal. A one-way ANOVA test can be used to assess if the residuals differ significantly. The null hypothesis \(H_0\) asserts that \((\gamma_1,\gamma_2,\cdots,\gamma_p)\) are jointly zero, rejecting \(H_0\) indicates that \(X\) Granger-cause \(Y\). 

\subsection{Cross-Correlation}\label{sec:cc}
Cross-correlation is a fundamental tool in time series analysis used to examine the linear relationship between two stochastic processes at different time lags. The cross-correlation function between two stationary time series $X_t$ and $Y_t$ is defined as:
\[
\rho_{XY}(k) = \frac{\mathbb{E}[(X_{t-k} - \mu_X)(Y_t - \mu_Y)]}{\sigma_X \sigma_Y},
\]
where $k$ represents the time lag, and $\mu_X, \mu_Y, \sigma_X, \sigma_Y$ denote the means and standard deviations of $X$ and $Y$, respectively. The cross-correlation function provides a quantitative measure of the strength and direction of the linear relationship between the two series at each lag $k$. \textcolor{black}{Significant cross-correlation at nonzero lags indicates temporal synchronization or lead-lag relationships but does not imply causal direction.} Cross-correlation analysis is frequently employed in econometrics and environmental sciences to explore temporal dependencies, particularly when evaluating synchronized behaviour or transmission effects between macroeconomic and environmental indicators. However, it is limited to linear dependencies and may not adequately capture nonlinear or nonstationary relationships \cite{box2015time}.

\subsection{Wavelet Coherence}\label{sec:wc}
Wavelet coherence is a powerful tool for examining localized correlation and phase relationships between two nonstationary time series across time and frequency domains. Unlike classical spectral approaches, wavelet coherence captures transient associations by decomposing the signals using continuous wavelet transforms. The continuous wavelet transform of a signal \(X(t)\) with respect to a mother wavelet \(\psi\) is defined as:
\[
W^X(a, b) = \int_{-\infty}^{\infty} X(t) \frac{1}{\sqrt{a}} \overline{\psi}\left(\frac{t - b}{a}\right) dt,
\]
where \(a\) is the scale (inversely proportional to frequency), \(b\) is the translational value, and \(^-\) represents the operation of complex conjugation. The cross-wavelet transform between two signals \(X(t)\) and \(Y(t)\) is given by:
\[
W^{XY}(a, b) = W^X(a, b) \; W^{\overline{Y}}(a, b).
\]
The wavelet coherence is then defined as:
\[
R^2(a, b) = \frac{|S(a^{-1}W^{XY}(a, b))|^2}{S(a^{-1}|W^X(a, b)|^2) \cdot S(a^{-1}|W^Y(a, b)|^2)},
\]
where \(S\) is a smoothing operator in time and scale. The resulting measure \(R^2(a,b) \in [0, 1]\) reflects the local linear correlation between \(X\) and \(Y\) at each time-scale location \((a, b)\) \cite{grinsted2004application}. Wavelet coherence is especially suitable for analyzing nonstationary and multiscale relationships in economic and environmental time series, where the interactions may vary across frequencies and evolve over time \cite{vacha2012co, ruehlemann2020wavelet}.

\subsection{Results of Association Analyses}\label{causal_ana_res}
\begin{figure}[h!]
    \centering
    \includegraphics[width=0.9\textwidth]{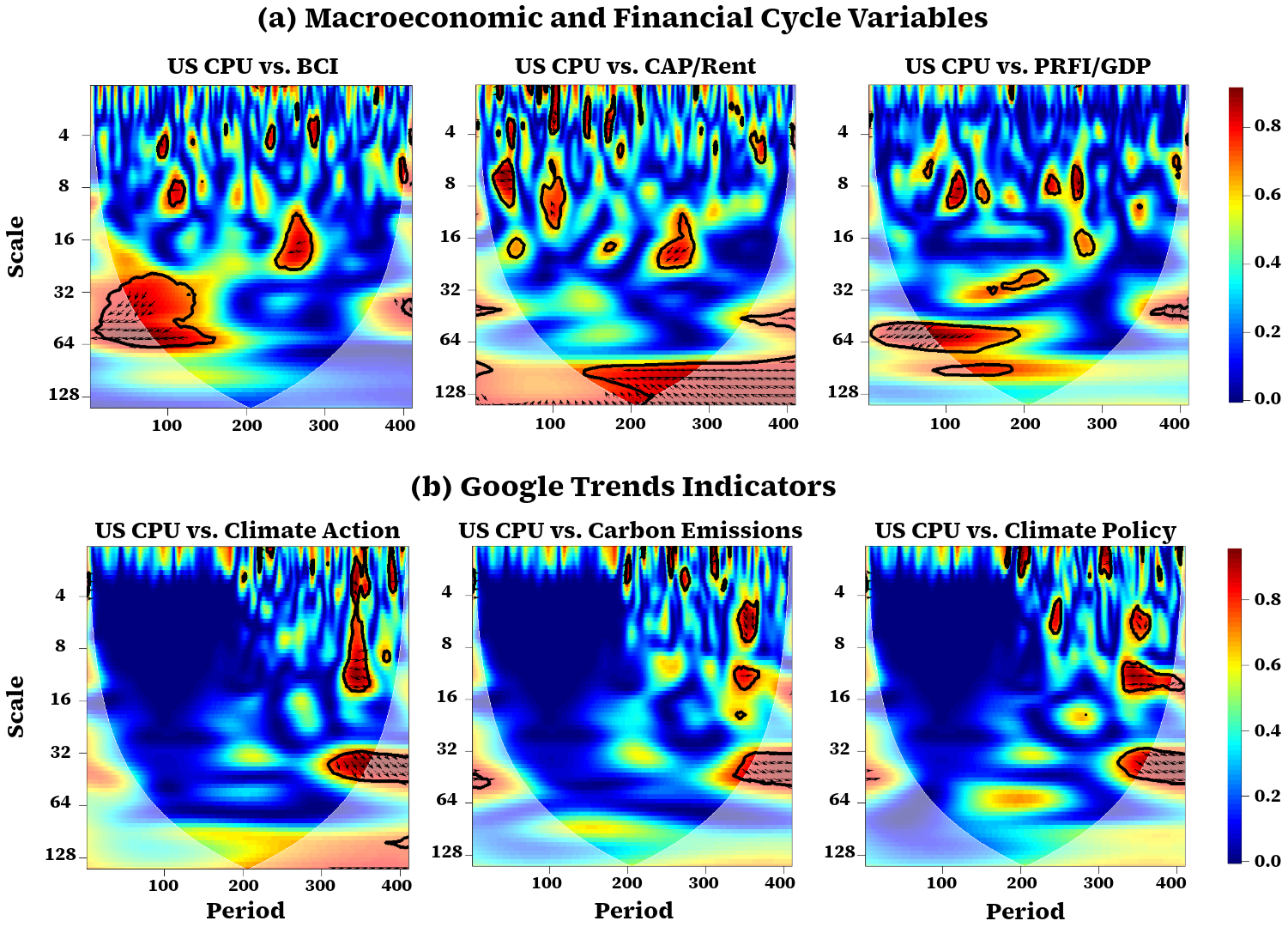}
    \caption{\textcolor{black}{Wavelet coherence plots showing time-frequency co-movement between the primary US CPU index and (a) macroeconomic and financial cycle variables and (b) Google Trends indicators. Color intensity represents the strength of association, with red indicating strong co-movement and blue weak co-movement. Arrows indicate phase relations, showing the lead-lag direction and whether series move in the same or opposite direction.}}
    \label{fig:Wavelet}
\end{figure}

\textcolor{black}{We apply the four complementary screening approaches to assess the predictive associations between the primary CPU index and 137 macroeconomic and financial cycle variables, as well as the Google Trends indicators.} Tables \ref{tab:causal_analyses} and \ref{tab:causal_analyses_google} summarize the results, where ``Y" indicates a statistically significant association detected by the screening procedure, while ``N" denotes the absence of such an association. \textcolor{black}{For transparency, Fig.~\ref{fig:Wavelet} provides a representative example of the wavelet coherence plots resulting from this screening method. Wavelet coherence plots are interpreted by considering both time and frequency simultaneously, represented on the horizontal and vertical axes, respectively. The color intensity indicates the degree of co-movement between the two series, with warmer colors corresponding to stronger association and cooler colors to weaker association. The orientation of the arrows shows the lead-lag relationship: arrows pointing up-left or down-right indicate that the first series leads the second, whereas arrows pointing up-right or down-left show that the second series leads the first. Regions outside the cone of influence correspond to areas where the relationship is not statistically significant.}

\vspace{0.5cm}

\begingroup
\footnotesize
\begin{longtable}{llllll}
\caption{Summary of statistical screening results for the macroeconomic and financial cycle variables.} \\
\toprule
Variables & Description & TE & GC & CC & W \\
\midrule
\endfirsthead

\toprule
\multicolumn{6}{c}{{\bfseries \tablename\ \thetable{} -- continued from previous page}} \\
\toprule
Variables & Full Form & TE & GC & CC & W \\
\midrule
\endhead

\midrule
\multicolumn{6}{r}{{Continued on next page}} \\
\endfoot

\bottomrule
\multicolumn{6}{l}{Note: TE = Transfer Entropy, GC = Granger Causality, CC = Cross-Correlation, W = Wavelet Coherence.} \\
\endlastfoot
AAA & Moody’s Seasoned Aaa Corporate Bond Yield & Y & N & N & N \\
AAAFFM & Moody’s Aaa Corporate Bond Minus FEDFUNDS & N & N & N & N \\
AMDMNOx & New Orders for Durable Goods & N & N & N & Y \\
AMDMUOx & Unfilled Orders for Durable Goods & N & N & N & Y \\
ANDENOx & New Orders for Nondefense Capital Goods & N & N & N & Y \\
AWHMAN  & New Orders for Nondefense Capital Goods & N & N & Y & Y \\
AWOTMAN & Average Weekly Overtime Hours : Manufacturing & N & N & N & Y \\
BAA & Moody’s Seasoned Baa Corporate Bond Yield & Y & N & N & Y \\
BAAFFM & Moody’s Baa Corporate Bond Minus FEDFUNDS & N & N & N & Y \\
BCI & Business Confidence Index & Y & N & Y & Y \\
BOGMBASE & Board of Governors Monetary Base & N & N & N & Y \\
BUSINVx & Total Business Inventories & Y & N & Y & Y \\
BUSLOANS & Commercial and Industrial Loans & N & Y & N & Y \\
CAPE & Cyclically Adjusted Price/Earnings ratio  & Y & N & Y & Y \\
CAPR & Cyclically Adjusted Price/Rent ratio & Y & Y & Y & Y \\
CE16OV & Civilian Employment & N & N & Y & Y \\
CES0600000007 & Average Weekly Hours: Goods-Producing & N & N & Y & Y \\
CES0600000008 & Average Hourly Earnings: Goods-Producing & N & Y & N & Y \\
CES1021000001 & All Employees: Mining and Logging: Mining & N & N & N & Y \\
CES2000000008 & Average Hourly Earnings: Construction & Y & Y & N & Y \\
CES3000000008 & Average Hourly Earnings: Manufacturing & N & N & N & Y \\
CLAIMSx & Initial Claims & N & Y & N & Y \\
CLF16OV & Civilian Labor Force Level & N & Y & N & Y \\
CLI & Composite Leading Indicator & N & N & Y & Y \\
CMRMTSPLx & Real Manufacturing and Trade Industries Sales & N & N & N & Y \\
COMPAPFFx & 3-Month Commercial Paper Minus FEDFUNDS & N & N & Y & Y \\
CONSPI & Nonrevolving consumer credit to Personal Income & N & Y & N & Y \\
CP3Mx & 3-Month AA Financial Commercial Paper Rate & N & N & Y & Y \\
CPIAPPSL & CPI : Apparel & N & N & N & Y \\
CPIAUCSL & CPI : All Items & N & N & N & Y \\
CPIMEDSL & CPI : Medical Care & N & N & N & Y \\
CPITRNSL & CPI : Transportation & N & Y & N & Y \\
CPIULFSL & CPI : All Items Less Food & N & N & N & Y \\
CRED & Credit/GDP ratio & N & Y & Y & Y \\
CRED\_GDP & Credit/GDP ratio & Y & Y & Y & Y \\
CUMFNS & Manufacturing capacity utilization & N & Y & Y & Y \\
CUUR0000SA0L2 & CPI : All items less shelter & N & N & N & Y \\
CUSR0000SA0L5 & CPI : All items less medical care & N & N & N & Y \\
CUSR0000SAC & CPI: Commodities & N & N & N & Y \\
CUUR0000SAD & CPI: Durables & Y & N & N & Y \\
CUSR0000SAS & CPI: Services & N & N & N & Y \\
DDURRG3M086SBEA & Personal Consumption Expenditures: Durable Goods & Y & Y & N & Y \\
DMANEMP & All Employees: Durable goods & N & Y & Y & Y \\
DNDGRG3M086SBEA & Personal Consumption Expenditures: Nondurable Goods & N & N & N & Y \\
DPCERA3M086SBEA & Real Personal Consumption Expenditure & Y & N & N & Y \\
DSERRG3M086SBEA & Personal Consumption Expenditures: Services & N & N & N & Y \\
DTCOLNVHFNM & Consumer Motor Vehicle Loans Outstanding & N & Y & N & Y \\
DTCTHFNM & Total Consumer Loans and Leases Outstanding & N & Y & N & N \\
EXCAUSx & Canada/US Foreign Exchange Rate & N & N & N & Y \\
EXJPUSx & Japan/US Foreign Exchange Rate & N & N & N & N \\
EXSZUSx & Switzerland/US Foreign Exchange Rate & N & N & N & Y \\
EXUSUKx & US/UK Foreign Exchange Rate & N & N & N & N \\
FEDFUNDS & Effective Federal Funds Rate & N & N & Y & Y \\
GS1 & 1-Year Treasury Rate & Y & N & Y & Y \\
GS5 & 5-Year Treasury Rate & N & N & N & Y \\
GS10 & 10-Year Treasury Rate & N & N & N & Y \\
HOUST & Housing Starts: Total New Privately Owned  & Y & N & N & Y \\
HOUSTMW & Housing Starts, Midwest & N & N & Y & Y \\
HOUSTNE & Housing Starts, Northeast & N & Y & Y & Y \\
HOUSTS & Housing Starts, South & Y & Y & N & Y \\
HOUSTW & Housing Starts, West & Y & N & N & Y \\
HPI & House Price Index & Y & N & Y & Y \\
HWI & Help-Wanted Index for US & N & Y & Y & Y \\
HWIURATIO & Ratio of Help Wanted/Number of Unemployed & Y & Y & Y & Y \\
INDPRO & Industrial Production (IP) Index  & N & Y & N & Y \\
INVEST & Securities in Bank Credit at All Commercial Banks & N & N & N & N \\
IPB51222S & IP: Residential Utilities & N & Y & N & Y \\
IPBUSEQ & IP: Business Equipment & N & N & N & Y \\
IPCONGD & IP: Consumer Goods & N & N & N & Y \\
IPDCONGD & IP: Durable Consumer Goods & N & N & N & Y \\
IPDMAT & IP: Durable Materials & N & N & N & Y \\
IPFINAL & IP: Final Products (Market Group) & N & N & N & Y \\
IPFPNSS & IP: Final Products and Nonindustrial Supplies & N & N & N & Y \\
IPFUELS & IP: Fuels & N & N & N & Y \\
IPMANSICS & IP: Manufacturing (SIC) & N & Y & N & Y \\
IPMAT & IP: Materials & N & Y & N & Y \\
IPNCONGD & IP: Nondurable Consumer Goods & N & N & N & Y \\
IPNMAT & IP: Nondurable Materials & Y & Y & N & Y \\
ISRATIOx & Total Business: Inventories to Sales Ratio & N & N & N & Y \\
M1SL & M1 Money Stock & N & Y & N & Y \\
M2REAL & Real M2 Money Stock & Y & N & Y & Y \\
M2SL & M2 Money Stock & N & N & N & Y \\
MANEMP & All Employees: Manufacturing & Y & N & Y & Y \\
Mortg & Household Real Mortgage Debt Growth & N & N & N & Y \\
Mortg/Income & Household Mortgage/Income ratio & N & Y & N & Y \\
NDMANEMP & All Employees: Nondurable goods & Y & N & Y & Y \\
NFCI & Chicago Fed national financial condition index & N & Y & N & Y \\
NONBORRES & Reserves Of Depository Institutions & N & N & N & Y \\
NONREVSL & Total Nonrevolving Credit & N & Y & N & Y \\
OILPRICEx & Crude Oil, spliced WTI and Cushing & N & N & N & Y \\
PAYEMS & All Employees: Total nonfarm & Y & N & Y & Y \\
PCEPI & Personal Consumption Expenditures: Chain Index & N & N & N & Y \\
PERMIT & New Private Housing Permits (SAAR) & Y & N & N & Y \\
PERMITMW & New Private Housing Permits, Midwest (SAAR) & N & N & N & Y \\
PERMITNE & New Private Housing Permits, Northeast (SAAR) & Y & N & Y & Y \\
PERMITS & New Private Housing Permits, South (SAAR) & N & N & Y & Y \\
PERMITW & New Private Housing Permits, West (SAAR) & N & N & N & Y \\
PIP/Income & Personal Interest Payments/Income ratio & Y & N & Y & Y \\
PPICMM & PPI: Metals and metal products: & N & N & N & Y \\
PRFI/GDP & Private Residential Fixed Investment/GDP ratio & N & Y & Y & Y \\
REALLN & Real Estate Loans at All Commercial Banks & N & N & N & Y \\
RETAILx & Retail and Food Services Sales & N & N & N & Y \\
RPI & Real Personal Income & N & Y & N & Y \\
S\&P.500 & S\&P’s Common Stock Price Index: Composite & N & N & N & Y \\
S\&P.div.yield & S\&P’s Composite Common Stock: Dividend Yield & N & N & N & Y \\
S\&P.PE.ratio & \footnotesize{S\&P’s Composite Common Stock: Price/Earnings Ratio} & N & N & N & Y \\
SP500 & Real S\&P500 Index Growth & N & N & N & Y \\
SRVPRD & All Employees: Service-Providing Industries & Y & Y & Y & Y \\
T10YFFM & 10-Year Treasury Constant Maturity Minus FEDFUNDS & N & N & Y & Y \\
T1YFFM & 1-Year Treasury Constant Maturity Minus FEDFUNDS & Y & N & Y & Y \\
T5YFFM & 5-Year Treasury Constant Maturity Minus FEDFUNDS  & N & N & Y & Y \\
TB3MS & 3-Month Treasury Bill Secondary Market Rate & N & N & Y & Y \\
TB3SMFFM & 3-Month Treasury Bill Minus FEDFUNDS & Y & N & Y & Y \\
TB6MS & 6-Month Treasury Bill Secondary Market Rate & Y & N & Y & Y \\
TB6SMFFM & 6-Month Treasury Bill Minus FEDFUNDS & Y & N & Y & Y \\
TOTRESNS & Total Reserves of Depository Institutions & N & N & N & Y \\
UEMP15OV & Civilians Unemployed - 15 Weeks \& Over & Y & Y & N & Y \\
UEMP15T26 & Civilians Unemployed for 15-26 Weeks & Y & Y & N & Y \\
UEMP27OV & Civilians Unemployed for 27 Weeks \& Over & N & N & N & Y \\
UEMP5TO14 & Civilians Unemployed for 5-14 Weeks & N & Y & N & Y \\
UEMPLT5 & Civilians Unemployed for Less than 5 Weeks & N & N & N & Y \\
UEMPMEAN & Average Duration of Unemployment (Weeks) & Y & Y & N & Y \\
UMCSENTx & Consumer Sentiment Index & Y & N & N & Y \\
UNRATE & Civilian Unemployment Rate & Y & Y & Y & Y \\
USCONS & All Employees: Construction & N & N & N & Y \\
USFIRE & All Employees: Financial Activities & Y & N & N & Y \\
USGOOD & All Employees: Goods-Producing Industries & N & N & Y & Y \\
USGOVT & All Employees: Government & Y & N & N & Y \\
USTPU & All Employees: Trade, Transportation \& Utilities & Y & N & Y & Y \\
USTRADE & All Employees: Retail Trade & Y & N & Y & Y \\
USWTRADE & All Employees: Wholesale Trade & Y & N & Y & Y \\
VIXCLSx & CBOE Volatility Index & N & N & N & Y \\
W875RX1 & Real Personal Income Excluding Transfer Receipts & Y & Y & N & Y \\
WPSFD49207 & PPI by Commodity: Finished Goods series & N & Y & N & Y \\
WPSFD49502 & PPI by Commodity: Personal Consumption Goods & N & Y & N & Y \\
WPSID61 & PPI by Commodity Type: Processed Goods & N & N & N & Y \\
WPSID62 & PPI by Commodity Type: Unprocessed Goods & N & N & N & Y 
\label{tab:causal_analyses}
\end{longtable}
\endgroup

\vspace{0.5cm}

\begingroup
\footnotesize
\begin{longtable}{llllll}
\caption{Summary of statistical screening results for Google Trends indicators.} \\
\toprule
Search Terms & Description & TE & GC & CC & W \\
\midrule
\endfirsthead

\toprule
\multicolumn{6}{c}{{\bfseries \tablename\ \thetable{} -- continued from previous page}} \\
\toprule
Search Terms & Description & TE & GC & CC & W \\
\midrule
\endhead

\midrule
\multicolumn{6}{r}{{Continued on next page}} \\
\endfoot

\bottomrule
\multicolumn{6}{l}{Note: TE = Transfer Entropy, GC = Granger Causality, CC = Cross-Correlation, W = Wavelet Coherence.} \\
\endlastfoot

Carbon Credits (Worldwide) & Tradable credits to emit CO\textsubscript{2}. & Y & N & Y & N \\
Carbon Emissions (Worldwide) & Total release of CO\textsubscript{2} into atmosphere. & Y & Y & Y & Y \\
Carbon Footprint (Worldwide) & Individual or firm emissions. & Y & N & Y & Y \\
Carbon Tax (Worldwide) & Taxes on carbon emissions. & Y & Y & Y & Y \\
Clean Energy (Worldwide) & Energy from renewable sources. & Y & Y & Y & Y \\
Climate Action (Worldwide) & Measures to mitigate climate change. & Y & Y & Y & Y \\
Climate News (Worldwide) & Media coverage on climate change. & Y & Y & Y & Y \\
Climate Policy (Worldwide) & Policies addressing climate change. & Y & Y & Y & Y \\
Climate Risk (Worldwide) & Negative effects from climate change. & Y & Y & Y & Y \\
Climate Technology (Worldwide) & Technology reducing GHG emissions. & Y & Y & Y & Y \\
Electric Vehicle (Worldwide) & Battery-powered vehicles. & Y & N & Y & Y \\
Energy Efficiency (Worldwide) & Using less energy for same results. & Y & N & Y & Y \\
Energy Policy (Worldwide) & Regulations on energy consumption. & Y & Y & Y & Y \\
Energy Transition (Worldwide) & Shift to renewable sources. & Y & Y & Y & Y \\
Environmental Policy (Worldwide) & Policies addressing environmental issues. & Y & N & Y & Y \\
Environmental Tax (Worldwide) & Taxes on harmful activities. & Y & Y & Y & Y \\
Global Warming Policy (Worldwide) & Policies to limit temperature rise. & Y & N & Y & Y \\
Green Finance (Worldwide) & Investments promoting sustainability. & Y & N & Y & Y \\
Green Infrastructure (Worldwide) & Sustainable urban planning. & Y & Y & Y & Y \\
Green Jobs (Worldwide) & Employment in sustainable sectors. & Y & N & Y & Y \\
Green Technology (Worldwide) & Sustainable, eco-friendly technology. & Y & N & Y & Y \\
Greenhouse Gas Emissions (Worldwide) & Emissions of CO\textsubscript{2}, CH\textsubscript{4}, and N\textsubscript{2}O. & Y & Y & Y & Y \\
Greenwashing (Worldwide) & False sustainability claims. & Y & Y & Y & Y \\
Renewable Energy (Worldwide) & Energy from natural sources. & Y & N & Y & Y \\
Sustainability (Worldwide) & Using resources wisely. & Y & Y & Y & Y \\
 Sustainable Development (Worldwide) & Economic growth with sustainability. & Y & Y & Y & Y \\
{\footnotesize Sustainable Development Goals (Worldwide)} & UN-adopted framework of 17 goals. & Y & N & Y & Y \\
Sustainable Living (Worldwide) & Lifestyle reducing natural resources. & Y & N & Y & Y \\
UN Climate Conference (Worldwide) & Annual international climate summit. & Y & N & Y & Y \\
Zero Emissions (Worldwide) & Completely eliminating GHG emissions. & Y & Y & Y & Y 
\label{tab:causal_analyses_google}
\end{longtable}
\endgroup

\section{Components of the BSTS Model}\label{bsts_components}

The BSTS framework allows modeling various components to capture different features of a time series. Commonly used components include:
\begin{enumerate}
    \item Local Level: This component models the series level as a random walk, allowing it to evolve gradually and unpredictably over time by adding small random changes at each step:
    \[
    \mu_t = \mu_{t-1} + u_t, \quad u_t \sim \mathcal{N}(0, \sigma_u^2).
    \]
    
    \item Local Linear Trend: This extends the local level by adding a slope term \( \delta_t \) to account for trends that can change smoothly:
    \[
    \mu_t = \mu_{t-1} + \delta_{t-1} + u_t, \quad \delta_t = \delta_{t-1} + v_t, \quad v_t \sim \mathcal{N}(0, \sigma_v^2).
    \]

    \item Autoregressive Process: This component introduces an autoregressive structure of order \( p \), denoted AR(\(p\)), for some positive integer \( p \). In an AR(\(p\)) process, the current state depends on its previous \( p \) lagged observations:
  \[
    \mu_t = \sum_{i=1}^{p} \phi_i \mu_{t-i} + \tilde{u}_t, \quad \tilde{u}_t \sim \mathcal{N}(0, \sigma_{\tilde{u}}^2).
\]

    \item Seasonal: This component captures recurring seasonal patterns by including a seasonal effect \( \tau_t \) modeled through dummy variables. The parameter \( S \) denotes the length of the seasonal cycle. The seasonal component evolves according to the equation below, which imposes the constraint that seasonal effects sum to zero over one full cycle of length \( S \).
    \[
    \tau_t = -\sum_{s=1}^{S-1} \tau_{t-s} + w_t, \quad w_t \sim \mathcal{N}(0, \sigma_w^2).
    \]
    
    \item Regression: This component allows the model to include external covariates \( x_t \) with coefficients \( \beta \). The model employs a spike-and-slab prior to select relevant predictors by shrinking the coefficients of irrelevant variables toward zero with high probability. The full observation equation is:
    \[
    y_t = \mu_t + \tau_t + \beta^\top x_t + \epsilon_t,
    \]
    where \( \epsilon_t \sim \mathcal{N}(0, \sigma^2) \) is the observation noise.
\end{enumerate}
By combining these components, the BSTS model can flexibly capture a wide range of patterns, including gradual shifts, periodic fluctuations, and the influence of external covariates, found in macroeconomic time series.

\section{Empirical Setup and Results}\label{appendix_emp}
\textcolor{black}{In this section, we introduce the baseline models and report the forecasting results for the alternative US CPU index and the Global CPU index.}

\subsection{Baseline Models}\label{appendix_baseline}
The performance of the BSTS model was evaluated against several statistical, machine learning, and deep learning models. The comparison was conducted using the following competitive frameworks:

\begin{itemize}

\item \textit{Autoregressive Integrated Moving Average with exogenous variables} (ARIMA-X) model captures linear dependencies by integrating: an autoregressive (AR) component, differencing (I), and a moving average (MA) component \cite{box1970}. In this framework, initially, differencing of order $d$ is applied to obtain a stationary series. Then, the AR component models $p$-lagged values of the series, while the MA component models $q$-lagged residuals. The coefficients are estimated by minimizing the Akaike Information Criterion (AIC).

\item \textit{Autoregressive Fractionally Integrated Moving Average with exogenous variables} (ARFIMA-X) model extends the traditional ARIMA model by allowing the differencing parameter $d$ to range between $(0, 0.5)$, enabling ARFIMA$(p,d,q)$ to effectively handle time series exhibiting long-range dependencies, where correlations decay slowly over time \cite{granger1980}.

\item \textit{Autoregressive Neural Network with exogenous variables} (ARNN-X) model generalizes the feed-forward neural network structure to handle autoregressive time series processes \cite{faraway1998}. In the ARNN$(p, k)$ model, $p-$lagged values are fed into the input layer and processed by $k$ neurons in the hidden layer. Commonly, $k$ is set to $ \lfloor\frac{p + 1}{2}\rfloor$. The model is first initialized with random values and trained using the gradient descent back-propagation algorithm \cite{rumelhart1986}; thus achieving robust learning and preventing overfitting \cite{hyndman2018forecasting}.

\item \textit{Neural Basis Expansion Analysis for Time Series with exogenous variables} (NBeats-X) framework consists of multiple blocks, each comprised of two main layers \cite{oreshkin2019}. The first layer captures and models the general patterns in the data, while the second layer refines the forecasts obtained from the first layer by re-modeling the residuals. This iterative approach improves the overall forecasting accuracy.

\item \textit{Neural Hierarchical Interpolation for Time Series with exogenous variables} (NHiTS-X) is an extension of the NBeats framework by imposing a hierarchical structure. Similar to NBeats, NHiTS consists of a sequence of blocks, albeit with a hierarchical design to better capture complex dependencies and relationships in the data \cite{challu2023}.

\item \textit{Decomposition-based Linear model with exogenous variables} (DLinear-X) decomposes the input time series into trend and seasonal components via moving average. Each component then goes through an independent linear layer, and the sum of their results is the final forecast. This method provides better performance by dealing with the trends in the data \cite{zeng2023}.

\item \textit{Normalization-based Linear model with exogenous variables} (NLinear-X) normalizes the input time series by subtracting its last value before passing it through a linear layer; then, the processed output has the subtracted value added back to produce the final forecast. This normalization technique is specially devised to cope with distribution shifts between the training and testing datasets \cite{zeng2023}.
\end{itemize}

\subsection{Empirical Results}\label{appendix_emp_res}
\textcolor{black}{We report the forecast performance evaluation for the alternative US and Global CPU indices. Tables~\ref{tab:performance_metrics_XMG_AUS} and~\ref{tab:performance_metrics_XM_AUS} present the results for the alternative US CPU index using covariate sets X$_\text{MG}$ and X$_\text{M}$, respectively, while Tables~\ref{tab:performance_metrics_XMG_Global} and~\ref{tab:performance_metrics_XM_Global} show the corresponding results for the Global CPU index.}

\begin{table*}[!ht]
\caption{\textcolor{black}{Evaluation of the BSTS-X$_\text{MG}$ model's performance relative to baselines across all forecast horizons for the alternative US CPU index using macro-financial variables and Google Trends indicators. The \textbf{\underline{best}} and \textbf{\textit{second-best}} results are highlighted.}}
\centering
\scriptsize
\begin{adjustbox}{width=1\textwidth, height= 3.5cm}
\begin{tabular}{cccccccccc}
\hline
Horizon & Metric & ARFIMA-X$_\text{MG}$ & ARIMA-X$_\text{MG}$ & ARNN-X$_\text{MG}$ & BSTS-X$_\text{MG}$ & NBEATS-X$_\text{MG}$ & NHiTS-X$_\text{MG}$ & DLinear-X$_\text{MG}$ & NLinear-X$_\text{MG}$
\\\hline

$h = 3$ & MAPE & 28.235 & \textbf{\textit{25.453}} & 74.700 & \textbf{\underline{23.856}} & 72.552 & 84.706 & 265.132 & 112.003 \\
        & SMAPE & 24.002 & \textbf{\underline{21.532}} & 47.242 & \textbf{\textit{22.791}} & 43.399 & 69.297 & 102.337 & 64.737 \\
        & MAE & 0.456 & \textbf{\underline{0.405}} & 1.201 & \textbf{\textit{0.428}} & 1.111 & 1.566 & 4.509 & 1.975 \\
        & MASE & 0.909 & \textbf{\underline{0.808}} & 2.393 & \textbf{\textit{0.854}} & 2.214 & 3.121 & 8.985 & 3.937 \\
        & RMSE & 0.540 & \textbf{\textit{0.482}} & 1.409 & \textbf{\underline{0.435}} & 1.444 & 1.579 & 4.879 & 2.225 \\
\hline
$h = 6$ & MAPE & 25.224 & \textbf{\underline{20.882}} & 61.151 & \textbf{\textit{23.627}} & 57.011 & 53.303 & 115.042 & 118.338 \\
        & SMAPE & 21.620 & \textbf{\underline{18.663}} & 40.595 & \textbf{\textit{23.954}} & 38.642 & 38.822 & 84.015 & 71.313 \\
        & MAE & \textbf{\textit{0.439}} & \textbf{\underline{0.375}} & 1.046 & 0.469 & 1.078 & 0.978 & 2.568 & 2.422 \\
        & MASE & \textbf{\textit{0.833}} & \textbf{\underline{0.711}} & 1.983 & 0.890 & 2.043 & 1.855 & 4.869 & 4.593 \\
        & RMSE & 0.519 & \textbf{\underline{0.472}} & 1.278 & \textbf{\textit{0.506}} & 1.350 & 1.157 & 3.003 & 2.650 \\
\hline
$h = 12$ & MAPE & \textbf{\textit{36.754}} & 39.545 & \textbf{\underline{27.542}} & 43.469 & 71.258 & 157.623 & 101.858 & 175.098 \\
         & SMAPE & 39.351 & \textbf{\textit{29.447}} & \textbf{\underline{22.865}} & 44.755 & 50.696 & 78.230 & 71.477 & 134.382 \\
         & MAE & 0.789 & \textbf{\textit{0.664}} & \textbf{\underline{0.488}} & 0.928 & 1.328 & 2.973 & 1.920 & 3.258 \\
         & MASE & 1.040 & \textbf{\textit{0.876}} & \textbf{\underline{0.643}} & 1.224 & 1.751 & 3.920 & 2.532 & 4.295 \\
         & RMSE & 0.892 & \textbf{\textit{0.815}} & \textbf{\underline{0.615}} & 1.140 & 1.500 & 3.309 & 2.314 & 3.652 \\
\hline
$h = 24$ & MAPE & 41.518 & \textbf{\textit{35.040}} & 55.712 & \textbf{\underline{33.092}} & 57.145 & 58.213 & 99.903 & 95.327 \\
         & SMAPE & 56.688 & \textbf{\textit{43.424}} & 80.431 & \textbf{\underline{36.085}} & 48.827 & 58.919 & 92.512 & 81.235 \\
         & MAE & 1.411 & \textbf{\textit{1.193}} & 1.708 & \textbf{\underline{1.028}} & 1.399 & 1.637 & 2.321 & 2.465 \\
         & MASE & 1.158 & \textbf{\textit{0.980}} & 1.402 & \textbf{\underline{0.844}} & 1.149 & 1.345 & 1.906 & 2.024 \\
         & RMSE & 1.841 & \textbf{\textit{1.613}} & 2.094 & \textbf{\underline{1.416}} & 1.749 & 2.104 & 2.830 & 3.095 \\
\hline
\end{tabular}
\label{tab:performance_metrics_XMG_AUS}
\end{adjustbox}
\end{table*}

\begin{table*}[!ht]
\caption{\textcolor{black}{Evaluation of the BSTS-X$_\text{M}$ model's performance relative to baselines across all forecast horizons for the alternative CPU index using macro-financial variables. The \textbf{\underline{best}} and \textbf{\textit{second-best}} results are highlighted.}}
\centering
\scriptsize
\begin{adjustbox}{width=\textwidth, height = 3.5cm}
\begin{tabular}{cccccccccc}
\hline
Horizon & Metric & ARFIMA-X$_\text{M}$ & ARIMA-X$_\text{M}$ & ARNN-X$_\text{M}$ & BSTSX$_\text{M}$ & NBEATS-X$_\text{M}$ & NHiTS-X$_\text{M}$ & DLinear-X$_\text{M}$ & NLinear-X$_\text{M}$ \\\hline

$h = 3$ & MAPE & 25.805 & 38.530 & 35.742 & \textbf{\textit{23.604}} & 47.997 & 59.063 & 57.061 & \textbf{\underline{19.440}} \\
 & SMAPE & 22.204 & 28.426 & 26.582 & \textbf{\underline{21.697}} & 33.872 & 40.144 & 51.832 & \textbf{\textit{22.104}} \\
 & MAE & \textbf{\textit{0.419}} & 0.609 & 0.547 & \textbf{\underline{0.407}} & 0.758 & 0.959 & 1.231 & \textbf{\underline{0.407}} \\
 & MASE & \textbf{\textit{0.834}} & 1.214 & 1.090 & \textbf{\underline{0.810}} & 1.510 & 1.911 & 2.452 & \textbf{\underline{0.810}} \\
 & RMSE & 0.480 & 0.780 & 0.707 & \textbf{\underline{0.428}} & 0.935 & 1.138 & 1.467 & \textbf{\textit{0.475}} \\
\hline
$h = 6$ & MAPE & \textbf{\underline{20.882}} & 33.645 & 36.804 & \textbf{\textit{30.750}} & 51.802 & 56.481 & 98.123 & 36.783 \\
 & SMAPE & \textbf{\underline{18.663}} & \textbf{\textit{26.258}} & 27.662 & 32.924 & 38.965 & 53.797 & 122.830 & 52.845 \\
 & MAE & \textbf{\underline{0.375}} & \textbf{\textit{0.581}} & 0.620 & 0.630 & 1.024 & 1.018 & 2.004 & 0.721 \\
 & MASE & \textbf{\underline{0.711}} & \textbf{\textit{1.103}} & 1.176 & 1.195 & 1.942 & 1.930 & 3.800 & 1.368 \\
 & RMSE & \textbf{\underline{0.472}} & 0.726 & 0.772 &\textbf{\textit{0.689}} & 1.158 & 1.099 & 2.123 & 0.895 \\
\hline
$h = 12$ & MAPE & \textbf{\underline{27.542}} & 46.815 & \textbf{\textit{39.933}} & 47.786 & 76.817 & 63.965 & 128.603 & 94.270 \\
 & SMAPE & \textbf{\underline{22.865}} & 59.220 & 47.322 & \textbf{\textit{34.600}} & 51.799 & 41.271 & 102.699 & 86.979 \\
 & MAE & \textbf{\underline{0.488}} & 1.064 & 0.886 & \textbf{\textit{0.820}} & 1.434 & 1.124 & 2.546 & 1.740 \\
 & MASE & \textbf{\underline{0.643}} & 1.403 & 1.168 & \textbf{\textit{1.081}} & 1.890 & 1.482 & 3.357 & 2.295 \\
 & RMSE & \textbf{\underline{0.615}} & 1.276 & 1.128 & \textbf{\textit{0.964}} & 1.579 & 1.444 & 3.241 & 2.185 \\
\hline
$h = 24$ & MAPE & \textbf{\textit{35.036}} & 46.852 & 39.040 & \textbf{\underline{34.378}} & 49.729 & 89.399 & 112.513 & 111.024 \\
 & SMAPE & \textbf{\textit{43.120}} & 60.537 & 48.452 & \textbf{\underline{41.070}} & 45.857 & 59.336 & 115.905 & 76.336 \\
 & MAE & \textbf{\textit{1.186}} & 1.430 & 1.224 & \textbf{\underline{1.141}} & 1.331 & 1.975 & 2.922 & 2.629 \\
 & MASE & \textbf{\textit{0.974}} & 1.174 & 1.005 & \textbf{\underline{0.937}} & 1.093 & 1.622 & 2.400 & 2.159 \\
 & RMSE & \textbf{\textit{1.602}} & 1.788 & 1.606 & \textbf{\underline{1.538}} & 1.667 & 2.472 & 3.379 & 3.231 \\
\hline
\end{tabular}
\label{tab:performance_metrics_XM_AUS}
\end{adjustbox}
\end{table*}

\begin{table*}[!ht]
\caption{\textcolor{black}{Evaluation of the BSTS-X$_\text{MG}$ model's performance relative to baselines across all forecast horizons for the Global CPU index using macro-financial variables and Google Trends indicators. The \textbf{\underline{best}} and \textbf{\textit{second-best}} results are highlighted.}}
\centering
\scriptsize
\begin{adjustbox}{width=1\textwidth, height= 3.5cm}
\begin{tabular}{cccccccccc}
\hline
Horizon & Metric & ARFIMA & ARIMA & ARNN-X$_\text{MG}$ & BSTS-X$_\text{MG}$ & NBEATS-X$_\text{MG}$ & NHiTS-X$_\text{MG}$ & DLinear-X$_\text{MG}$ & NLinear-X$_\text{MG}$
\\\hline

$h = 3$ & MAPE & \textbf{\textit{21.678}} & 25.523 & 21.820 & \textbf{\underline{18.107}} & 72.636 & 47.077 & 60.955 & 44.037 \\
        & SMAPE & \textbf{\textit{19.034}} & 22.124 & 19.362 & \textbf{\underline{16.070}} & 53.041 & 37.497 & 46.696 & 35.986 \\
        & MAE & \textbf{\textit{31.623}} & 37.576 & 32.322 & \textbf{\underline{26.115}} & 111.029 & 70.564 & 94.070 & 67.680 \\
        & MASE & \textbf{\textit{1.223}} & 1.454 & 1.250 & \textbf{\underline{1.010}} & 4.295 & 2.730 & 3.639 & 2.618 \\
        & RMSE & 35.179 & 40.432 & \textbf{\textit{34.128}} & \textbf{\underline{30.484}} & 111.430 & 72.594 & 94.266 & 68.074 \\
\hline
$h = 6$ & MAPE & 16.361 & 16.517 & \textbf{\textit{16.314}} & \textbf{\underline{16.176}} & 49.647 & 34.694 & 40.944 & 39.588 \\
        & SMAPE & 18.174 & 17.444 & \textbf{\textit{17.251}} & \textbf{\underline{16.179}} & 38.604 & 29.221 & 32.472 & 32.322 \\
        & MAE & 32.047 & 30.796 & \textbf{\textit{30.499}} & \textbf{\underline{28.673}} & 84.072 & 57.717 & 67.496 & 65.293 \\
        & MASE & 1.145 & 1.100 & \textbf{\textit{1.090}} & \textbf{\underline{1.024}} & 3.004 & 2.062 & 2.411 & 2.333 \\
        & RMSE & 38.664 & 34.548 & \textbf{\textit{34.421}} & \textbf{\underline{32.601}} & 88.740 & 62.155 & 74.414 & 74.278 \\
\hline
$h = 12$ & MAPE & \textbf{\textit{12.979}} & 16.409 & \textbf{\underline{11.397}} & 15.553 & 73.687 & 43.982 & 40.884 & 39.604 \\
         & SMAPE & \textbf{\textit{13.921}} & 15.176 & \textbf{\underline{12.371}} & 14.596 & 51.559 & 34.386 & 31.838 & 30.210 \\
         & MAE & \textbf{\textit{23.595}} & 26.080 & \textbf{\underline{21.095}} & 25.012 & 116.900 & 69.070 & 64.398 & 62.770 \\
         & MASE & \textbf{\textit{1.114}} & 1.231 & \textbf{\underline{0.996}} & 1.181 & 5.518 & 3.260 & 3.040 & 2.963 \\
         & RMSE & 32.527 & 29.357 & \textbf{\textit{28.467}} & \textbf{\underline{28.018}} & 125.711 & 77.471 & 75.484 & 78.708 \\
\hline
$h = 24$ & MAPE & 29.670 & \textbf{\textit{22.647}} & 23.918 & \textbf{\underline{18.247}} & 33.289 & 31.992 & 41.402 & 34.669 \\
         & SMAPE & 35.654 & \textbf{\textit{26.065}} & 27.752 & \textbf{\underline{20.305}} & 29.736 & 28.029 & 40.439 & 34.690 \\
         & MAE & 56.315 & \textbf{\textit{43.536}} & 45.537 & \textbf{\underline{35.001}} & 54.837 & 52.483 & 71.323 & 60.076 \\
         & MASE & 1.464 & \textbf{\textit{1.132}} & 1.184 & \textbf{\underline{0.910}} & 1.426 & 1.365 & 1.854 & 1.562 \\
         & RMSE & 64.629 & \textbf{\textit{53.344}} & 53.903 & \textbf{\underline{44.656}} & 62.984 & 61.232 & 85.001 & 70.294 \\
\hline
\end{tabular}
\label{tab:performance_metrics_XMG_Global}
\end{adjustbox}
\end{table*}

\begin{table*}[!ht]
\caption{\textcolor{black}{Evaluation of the BSTS-X$_\text{M}$ model's performance relative to baselines across all forecast horizons for the Global CPU index using macro-financial variables. The \textbf{\underline{best}} and \textbf{\textit{second-best}} results are highlighted.}}
\centering
\scriptsize
\begin{adjustbox}{width=\textwidth, height = 3.5cm}
\begin{tabular}{cccccccccc}
\hline
Horizon & Metric & ARFIMA & ARIMA & ARNN-X$_\text{M}$ & BSTSX$_\text{M}$ & NBEATS-X$_\text{M}$ & NHiTS-X$_\text{M}$ & DLinear-X$_\text{M}$ & NLinear-X$_\text{M}$ \\\hline

$h = 3$ & MAPE & 21.678 & 25.523 & \textbf{\textit{11.617}} & 21.721 & 56.689 & 54.437 & 30.964 & \textbf{\underline{10.193}} \\
 & SMAPE & 19.034 & 22.124 & \textbf{\textit{10.850}} & 19.153 & 43.734 & 42.673 & 26.419 & \textbf{\underline{9.281}} \\
 & MAE & 31.623 & 37.576 & \textbf{\textit{17.083}} & 31.879 & 86.300 & 83.537 & 46.226 & \textbf{\underline{14.717}} \\
 & MASE & 1.223 & 1.454 & \textbf{\textit{0.661}} & 1.233 & 3.338 & 3.231 & 1.788 & \textbf{\underline{0.569}} \\
 & RMSE & 35.179 & 40.432 & \textbf{\underline{18.467}} & 34.705 & 87.543 & 83.834 & 47.823 & \textbf{\textit{19.896}} \\
\hline
$h = 6$ & MAPE & 16.361 & 16.517 & \textbf{\underline{14.931}} & \textbf{\textit{16.357}} & 43.532 & 28.335 & 15.792 & 23.141 \\
 & SMAPE & 18.174 & 17.444 & \textbf{\textit{16.544}} & 17.353 & 34.591 & 26.207 & \textbf{\underline{15.226}} & 27.960 \\
 & MAE & 32.047 & 30.796 & \textbf{\textit{29.458}} & 30.643 & 72.948 & 48.570 & \textbf{\underline{26.854}} & 46.196 \\
 & MASE & 1.145 & 1.100 & \textbf{\textit{1.052}} & 1.095 & 2.606 & 1.735 & \textbf{\underline{0.959}} & 1.650 \\
 & RMSE & 38.664 & \textbf{\underline{34.548}} & 35.911 & \textbf{\textit{34.564}} & 78.107 & 54.100 & 35.289 & 58.055 \\
\hline
$h = 12$ & MAPE & \textbf{\underline{12.979}} & 16.409 & 18.982 & \textbf{\textit{15.442}} & 87.385 & 47.419 & 30.174 & 17.991 \\
 & SMAPE & \textbf{\underline{13.921}} & 15.176 & 21.546 & \textbf{\textit{14.499}} & 58.467 & 37.072 & 25.068 & 19.554 \\
 & MAE & \textbf{\underline{23.595}} & 26.080 & 33.537 & \textbf{\textit{24.837}} & 139.414 & 75.537 & 47.434 & 31.696 \\
 & MASE & \textbf{\underline{1.114}} & 1.231 & 1.583 & \textbf{\textit{1.172}} & 6.581 & 3.566 & 2.239 & 1.496 \\
 & RMSE & 32.527 & \textbf{\textit{29.357}} & 39.140 & \textbf{\underline{27.913}} & 147.778 & 81.757 & 55.299 & 40.667 \\
\hline
$h = 24$ & MAPE & 29.670 & \textbf{\textit{22.647}} & 35.113 & \textbf{\underline{18.286}} & 30.848 & 30.722 & 31.056 & 31.936 \\
 & SMAPE & 35.654 & \textbf{\textit{26.065}} & 44.348 & \textbf{\underline{20.402}} & 32.174 & 29.215 & 29.843 & 38.486 \\
 & MAE & 56.315 & \textbf{\textit{43.536}} & 64.709 & \textbf{\underline{35.155}} & 55.289 & 52.665 & 53.416 & 60.074 \\
 & MASE & 1.464 & \textbf{\textit{1.132}} & 1.682 & \textbf{\underline{0.914}} & 1.438 & 1.369 & 1.389 & 1.562 \\
 & RMSE & 64.629 & \textbf{\textit{53.344}} & 72.359 & \textbf{\underline{44.900}} & 66.888 & 60.151 & 65.973 & 75.612 \\
\hline
\end{tabular}
\label{tab:performance_metrics_XM_Global}
\end{adjustbox}
\end{table*}

\end{document}